\newcommand{\version}{February 18, 2021}
\documentclass[a4paper,oneside,11pt,pdftex]{article}
\pdfoutput=1
\usepackage[bindingoffset=0.0cm,textheight=22.6cm,hdivide={*,16.0cm,*}, vdivide={*,22cm,*}]{geometry}
\usepackage{amsmath,amsfonts,amssymb}
\usepackage{mathtools}
\usepackage{empheq}	%% package for boxing multiline equations
\usepackage[pdftex]{graphicx}
\usepackage[T1]{fontenc}
\usepackage[utf8]{inputenc}
\usepackage{lmodern}
\usepackage{tensind}
\tensordelimiter{?}
\IfFileExists{dsfont.sty}
	{\usepackage{dsfont}
         \let\mathbb=\mathds
         \newcommand{\id}{\mathds{1}}}
	{\typeout{Package dsfont.sty was not found, using alternative macros.}
         \let\mathds=\mathbb
         \newcommand{\id}{\mbox{1 \kern-.59em \textrm{l}}}}

\usepackage[pdftex,hyperref,svgnames]{xcolor}
\usepackage[pdftex,bookmarksnumbered=true,breaklinks=true,colorlinks=true,linktocpage=true,linkcolor=MediumBlue,citecolor=ForestGreen,urlcolor=DarkRed]{hyperref}
\makeatletter
\AtBeginDocument{
  \hypersetup{
    pdftitle = {\@title},
    pdfauthor = {Daniel N. Blaschke and Fran\c{c}ois Gieres}
  }
}
\makeatother

\usepackage[numbers,square,comma,sort&compress]{natbib}

\usepackage{etoolbox}
\newtoggle{grouprefs}
\toggletrue{grouprefs}
\iftoggle{grouprefs}{
\newcommand{\GitmanProkhorov}{GitmanProkhorov}
\newcommand{\Garcia}{Garcia:1996ac}
\newcommand{\GarciaPerez}{GarciaPerez71}
\newcommand{\WittenEtal}{Witten:1986qs}
\newcommand{\Kijowski}{Kijowski81}
\newcommand{\becchi}{becchi}
\newcommand{\heinzlbakker}{heinzlbakker}
}
{
\newcommand{\GitmanProkhorov}{Gitman:1990,Prokhorov:2011}
\newcommand{\Garcia}{Garcia:1996ac,Garcia:1997vs}
\newcommand{\GarciaPerez}{GarciaPerez71,Garcia:1974}
\newcommand{\WittenEtal}{Witten:1986qs,Crnkovic:1986ex,Crnkovic:1987tz}
\newcommand{\Kijowski}{Kijowski81,Kijowski:1976ze}
\newcommand{\becchi}{Becchi:1974xu,Becchi:1974md,Becchi:1975nq}
\newcommand{\heinzlbakker}{Heinzl:2000ht,Bakker:2000mn}
}

\makeatletter

 \@addtoreset{equation}{section}
 \makeatother

\newcommand{\vp}{\varphi}
\newcommand{\mg}{\textbf{\texttt{g}}}
\newcommand{\mq}{q}

\newcommand{\pa}{\partial}

\newcommand{\varep}{\varepsilon}

\newcommand{\lie}{\mathtt{\mathbf{g}}}
\newcommand{\br}{\mathbb{R}}

\newcommand{\txt}[1]{\textrm{#1}}
\newcommand{\grad}{\overrightarrow{\textrm{grad}}}
\newcommand{\Div}{\textrm{div}}
\newcommand{\curl}{\overrightarrow{\textrm{curl}}}
\newcommand{\Tr}{\textrm{Tr}}

\newcommand{\eqnref}[1]{eqn.~(\ref{#1})}

\newcommand{\secref}[1]{section~\ref{#1}}

\newcommand{\ri}{\textrm{i}}
\newcommand{\re}{\textrm{e}}
\renewcommand{\th}{\theta}

\renewcommand{\l}{\lambda}
\newcommand{\m}{\mu}
\newcommand{\n}{\nu}
\newcommand{\nn}{\nonumber}

\newcommand{\pr}{\prime}

\newcommand{\sgn}[1]{\textrm{sgn}\!\left(#1\right)}

\newcommand{\Boxed}[1]{\setlength\fboxsep{0.4em}\boxed{\ #1 \ }}

\title{\texorpdfstring{\begin{flushright}
\vspace*{-2cm}{\small
LYCEN-2020-01
 \\[-1ex] LA-UR-20-21941}
       \end{flushright}\vspace{2em}}{}%
On the canonical formulation of gauge field theories \texorpdfstring{\\}{}and Poincar{\'e} transformations%
}

\date{\version}

\author{Daniel N. Blaschke\footnotemark[2]~
 and
 Fran\c{c}ois Gieres\footnotemark[3]}

\begin{document}
\renewcommand{\thepage}{\roman{page}}
%\today

% \graphicspath{{./}{./figures/}}

\maketitle
\thispagestyle{empty}
\begin{center}
\renewcommand{\thefootnote}{\fnsymbol{footnote}}
\vspace{-0.3cm}
\footnotemark[2]Los Alamos National Laboratory,\\Los Alamos, NM, 87545, (USA)\\[0.3cm]
\footnotemark[3]%
Institut de Physique des $2$ Infinis de Lyon, \\
Universit\'e de Lyon, Universit\'e Claude Bernard Lyon 1 and CNRS/IN2P3,\\Bat. P. Dirac, 4 rue Enrico Fermi,
F-69622-Villeurbanne (France)
\\[0.5cm]
\ttfamily{E-mail: dblaschke@lanl.gov, gieres@ipnl.in2p3.fr}
\end{center}

%\vspace{1.0em}
%\begin{center}
% \textsc{Dedicated to the memory of Manfred Schweda}
%\end{center}

\vspace{1.0em}
\begin{abstract}
We address the Hamiltonian formulation of classical gauge field theories
while putting forward
results some of which are not entirely new, though they
do not appear to be well known.
We refer in particular to the fact that neither
the \emph{canonical} energy momentum vector $(P^\m )$ nor the \emph{gauge invariant} energy momentum vector $(P_{\textrm{inv}} ^\m )$ do
generate space-time translations of the gauge field by means of the Poisson brackets:
In a general gauge, one has to consider the so-called kinematical energy momentum vector
and, in a specific gauge (like the radiation gauge in electrodynamics), one has to consider the Dirac brackets rather than the Poisson brackets.
Similar arguments apply to rotations and to Lorentz boosts and are of direct relevance
to the ``nucleon spin crisis'' since the spin of the proton involves a contribution which is due
to the angular momentum vector of gluons and thereby requires a proper treatment
of the latter.
We conclude with some comments on the relationships between the different approaches to quantization
(canonical quantization based on the classical Hamiltonian formulation,
Gupta-Bleuler, path integrals,
BRST, covariant canonical approaches).
% as well as axiomatic quantum field theory).

\end{abstract}

\newpage
 \renewcommand{\thepage}{\arabic{page}}
\setcounter{page}{0}
\tableofcontents
% \thispagestyle{empty}

% \newpage
% \renewcommand{\thepage}{\arabic{page}}
% \setcounter{page}{1}

%%%%%%%%%%%%%%%%%%%%%%%%%%%%
\section{Introduction}
%%%%%%%%%%%%%%%%%%%%%%%%%%%%

%A century ago
In 1918, Emmy Noether published her famous article on invariant variational problems
in which she stated and proved the so-called Noether theorem(s)~\cite{Noether:1918zz}. Over the years the latter have become a pillar of classical mechanics
and field theory, see reference~\cite{Kosmann} for an historical account and reference~\cite{Sundermeyer:2014kha} for a general discussion and various applications.
According to Noether's first theorem, the invariance of a field theoretic action functional $S[\phi ]$  under an $m$-dimensional Lie group (of global symmetry transformations)
implies the existence of $m$ local conservation laws for any solution $\phi$ of the equations of motion $\delta S/ \delta \phi =0$.
Following Noether's work, Felix Klein raised the question about the application of Noether's results to the free electromagnetic field.
In 1921, E.~Bessel-Hagen tackled this problem~\cite{Bessel-Hagen} while taking into account the remark made to him by E.~Noether that  the invariance
of the action functional $S[\phi ] \equiv \int_{\br ^n} d^nx \, {\cal L} (\phi, \pa_\mu \phi, x)$ allows for the addition of a total divergence
term $\pa_\mu \Omega^\mu (\phi,  x)$ to the Lagrangian density ${\cal L}$ (``divergence symmetry''):
By starting from the conformal invariance
of the free Maxwell equations (discovered in 1910 by H.~Bateman and E.~Cunningham) and cleverly combining with local gauge invariance (to which Noether's second theorem applies),
he could determine fifteen conserved, gauge invariant quantities in four dimensional Minkowski space.

Among the conformal transformations we have the Poincar\'e transformations and in particular the space-time translations.
More specifically, the invariance of the action under translations $x^\n \leadsto x^{\n} - a^\n$ in $\br^n$
implies the local conservation law for the canonical EMT (energy-momentum tensor), $\pa_\m T_{\textrm{can}} ^{\m \n}=0$, and thereby the existence of $n$ conserved ``charges''
$P^\n \equiv \int _{\br ^{n-1}} d^{n-1} x \, T^{0\n}_{\textrm{can}}$ which are interpreted as the total energy-momentum of the fields.
In their pioneering work on the ``quantum dynamics of wave fields'' of 1929~\cite{Heisenberg:1929xj},
W.~Heisenberg and W.~Pauli presented the general Lagrangian and Hamiltonian formulation of classical relativistic field theories
as well as the procedure of canonical quantization (based on equal-time commutation relations).
It is commonly believed that, within the Hamiltonian formulation of a classical field theory, the Noether charges
generate the symmetry transformations of the phase space variables $\vp\in \{ \phi, \pi \equiv \pa {\cal L}/ \pa \dot{\phi} \}$ by means of
the Poisson brackets, e.g. for infinitesimal translations, $\delta_{a} \vp (x)  \equiv \{\vp (x) , a^\m P_\m \} = a^\m \pa_\m \vp(x)$.
This is indeed the case for matter fields (scalar or Dirac fields), but, as we will discuss in detail in the present article,
it is definitely more subtle
for a gauge field $(A^\m )$:
This is due to the fact that the gauge invariance of the  action functional $S[A]$ implies the presence of constraints
for the phase space variables (as was already noted by Heisenberg and Pauli for electrodynamics in their pioneering work).

As was only realized recently~\cite{Salisbury:2017oev}, the  treatment of the constraints appearing in Lagrangian (or Hamiltonian) dynamical systems with
local symmetries like electrodynamics or general relativity has been systematically investigated in 1930 upon Pauli's impetus
by his assistant L\'eon Rosenfeld in a seminal work whose goal was the quantization of the
Maxwell-Dirac-Einstein field equations~\cite{Rosenfeld:2017umg}. In the sequel, Rosenfeld moved to other subjects and his work fell into oblivion.
In the late forties and fifties, P.~Bergmann and his collaborators~\cite{Salisbury:2012ona}
as well as P.A.M.~Dirac rediscovered the results found, or at least anticipated, twenty years earlier by L.~Rosenfeld
and they worked them out further (e.g. Dirac's modification of the Poisson brackets).
In particular, Dirac exposed the general approach to constrained Hamiltonian dynamics in his celebrated Yeshiva lectures
of 1964~\cite{Dirac:1964} (see~\cite{Hanson:1976cn,\GitmanProkhorov,Henneaux:1992,Wipf:1993} for more recent introductions).
%The treatment of constraints has been systematically investigated in the late forties by P.~Bergmann and his collaborators \cite{Salisbury:2012ona}
%and a general approach to constrained Hamiltonian dynamics has been put forward by P.A.M.~Dirac and exposed in his celebrated Yeshiva lectures
%of 1964~\cite{Dirac:1964} (see \cite{Hanson:1976cn,\GitmanProkhorov,Henneaux:1992,Wipf:1993} for more recent introductions).
 More recently, the quantization of (non-Abelian) gauge field theories
has been  revolutionized by the discovery  of the so-called BRST-symmetry (Becchi, Rouet, Stora 1974~\cite{\becchi}, Tyutin 1975~\cite{Tyutin:1975qk}) and its application to the
perturbative renormalization of these theories in their Lagrangian formulation.
Yet, the Hamiltonian formulation of classical Abelian or non-Abelian gauge field theory and the canonical approach  to its quantization
continue to represent a basic tool and useful device for exploring various aspects of gauge theories. Thus, it is worthwhile to have
a clear view of the action of Poincar\'e transformations on the phase space variables in the Hamiltonian formulation.
To a large extent, these aspects have already been addressed about forty years ago by some of the masters
of the subject (A.~J.~Hanson, T.~Regge and C.~Teitelboim) in their Roma lectures~\cite{Hanson:1976cn}.
The goal of the present article
(the impetus for which came in part from  our joint work
with M. Reboud and M.~Schweda~\cite{Blaschke:2016ohs}) is to give a short pedagogical account of these ideas.
%is to give a short pedagogical account of these ideas which is in part based on
%(and an extension of)  our joint work
%with M. Reboud and M.~Schweda~\cite{Blaschke:2016ohs}.
We hope that our presentation clarifies some misleading or erroneous statements made in the literature
and will prove to be useful as a complement to the basic textbook treatments of classical gauge theories and their quantization.

Our text is organized as follows.
In section \ref{sec:HamFormFT}, we briefly recall the definition and salient features of the Hamiltonian formulation of classical relativistic field theories.
As reviewed in  section~\ref{sec:freefields}, the description of the geometric transformations of matter fields
(scalar and Dirac fields) within this setting is unproblematic.
In section~\ref{sec:LagrForm}, we introduce the canonical and improved (gauge invariant) current densities and charges
following from the Poincar\'e invariance of the action functional for pure Yang-Mills theories.
The Hamiltonian formulation of Abelian and non-Abelian gauge field theories is then dealt with in section~\ref{sec:HamFormMaxwell}
and~\ref{sec:HamYM}, respectively.
The quantization procedure(s) for these theories are addressed in section~\ref{sec:AssessMax} and~\ref{sec:AssessYM}, respectively,
while the coupling to matter fields is considered in section~\ref{sec:matter}.
The identification of the physical observables of angular momentum (and its decomposition into
different contributions) is outlined in section~\ref{sec:obervables} while the concluding remarks
gather some remarks on other approaches to classical (gauge) field theories like
the multisymplectic or covariant phase space formulations.
In order to provide a better understanding of the structure of geometric symmetry transformations
in gauge theories and of their relationship with conserved quantities
we devote an appendix to a concise
and unified derivation of conserved gauge invariant currents associated to the conformal group.

\paragraph{Notation and conventions:}
We consider the natural system of units ($c\equiv 1 \equiv \hbar$ and $\varepsilon_0 = 1 \equiv \mu _0$ for electrodynamics).
Furthermore, we use the standard notation for the coordinates of $n$-dimensional Minkowski space,
i.e.
$x = (x^\m ) = (t, x^i) = (t, \vec x \, ) $ as well as the signature $(+, - , \cdots , - )$
for the Minkowski metric $\eta \equiv (\eta _{\m \n})$.

%After briefly reviewing some properties of free scalar and Dirac fields in section \ref{sec:freefields}, we proceed in section \ref{sec:LagrForm} with the Lagrangian formulation of pure gauge %theories.
%In section \ref{sec:HamFormMaxwell} we subsequently present an in-depth discussion of the Hamiltonian formulation of free Maxwell theory, whose quantization is remarked upon in section %\ref{sec:AssessMax}.
%Sections \ref{sec:HamYM} and \ref{sec:AssessYM} present the generalizations to non-Abelian gauge fields.
%After briefly commenting on the coupling to (scalar) matter in section \ref{sec:matter}, we close with reviewing some recent puzzles concerning the photon angular momentum in section %\ref{sec:obervables}.
%The concluding section finally gathers some remarks on other approaches.

%%%%%%%%%%%%%%%%%%%%%%%%%%%%%%%%%%%%%%%%%%%%%%%%%%%%%%%%%%%%%%%%%%%%%%%%%%%%%%
\section{Hamiltonian formulation of field theory}\label{sec:HamFormFT}

The Hamiltonian formulation of classical field theory in $\br^n$
is the starting point for its canonical quantization, and
we briefly recall~\cite{GreinerField}
 here its basics for a given Lagrangian density ${\cal L} (\phi, \pa_\m \phi )$.
The  \emph{canonical momentum} $\pi_\phi$  associated to the field $\phi$
is defined by $
\pi \equiv
\pi_\phi \equiv
\pa {\cal L} / \pa \dot{\phi}$
and the \emph{canonical Hamiltonian density} ${\cal H}$ is defined in terms of the fields $\phi$ and
their canonical momenta
$\pi$
 by means of a Legendre transformation:
\begin{align}
{\cal H}  \equiv \dot{\phi} \, \pi - {\cal L}
\, .
\label{eq:def-H}
\end{align}
Here, and in similar expressions to follow, the sum over all fields is
implicitly understood (e.g. the sum over $\phi$ and $\phi^*$ in the case of a complex scalar field $\phi$).

We note that in the simplest situation (which is realized for instance
for a free real scalar field),
the relation  $\pi \equiv \pa {\cal L} / \pa \dot{\phi}$ can be solved for $\dot{\phi}$ as a function of
$\pi$ (and possibly $\phi$ and/or the spatial derivatives $\pa_k \phi$).
The Hamiltonian function $H[\phi, \pi ] \equiv \int d^{n-1} x \, {\cal H}(\phi, \pi , \pa_k \phi)$
is now to be viewed as a functional of the fields $\phi$ and $\pi$.

For any two functionals  $F,G$ of bosonic fields  $\phi$ and $\pi$,
the \emph{canonical Poisson bracket} is defined at fixed time $t$ by
\begin{equation}
\label{eq:defpoisson}
\{ F  , G \} \equiv
\int d^{n-1}x \, \left(
\frac{\delta F }{ \delta \phi} \frac{\delta G}{\delta \pi}
- \frac{\delta F}{\delta \pi} \frac{\delta G}{\delta \phi}
\right)
\, .
\end{equation}
This bracket is bilinear and antisymmetric in its arguments, and it satisfies the  Jacobi identity as well as the Leibniz product rule
for each of its arguments, e.g. for the second argument: $\{ F , GH \}=\{ F  , G \} H + G \{ F , H \}$.
The general expression~\eqref{eq:defpoisson} yields the fundamental bracket of fields for any fixed time $t$,
e.g. for a single real scalar field $\phi$:
\begin{align}
\label{eq:fundPB}
\{ \phi (t, \vec x \, ) , \pi (t , \vec y \, ) \} & = \delta (\vec x - \vec y \, )
\,.
\end{align}

For a given Hamiltonian function $H[\phi, \pi ] \equiv \int d^{n-1} x \, {\cal H}(\phi, \pi , \pa_k \phi)$,
the \emph{time evolution} of a functional $F[\phi , \pi]$
is given by
\[
\dot{F} = \{ F, H \}
\,.
\]
For example
\[
\dot{\phi} = \{ \phi, H \}= \frac{\delta H }{\delta \pi} \, , \qquad \dot{\pi} = \{ \pi, H \}= -  \frac{\delta H}{\delta \phi} \, ,
\]
i.e. the Hamiltonian equations of motion of the field theoretic system described by $H[\phi , \pi]$.

Let us again consider the particular case of a free real scalar field $\phi$.
The \emph{space-time translations of the phase space variables} $\vp \equiv (\phi , \pi)$ are then generated by the conserved Noether charges $P^\m$
(which are associated to the translation invariance of the action)
and these transformations are described in terms of the canonical Poisson bracket:
For $a \equiv ( a^\m ) \in \br ^n$, we have
\begin{align}
 \Boxed{
\delta_{a} \vp (x)  \equiv \{\vp (x) , a^\m P_\m \} = a^\m \pa_\m \vp(x)
}
\, .
\label{eq:PgenTransl}
\end{align}
Similarly the transformation laws of the phase space variables under Lorentz transformations are generated by the Noether
charges $J^{\rho \sigma} = - J^{\sigma \rho}$ associated to the Lorentz invariance of the action:
Denoting the constant symmetry parameters by $\varepsilon_{\rho \sigma} = - \varepsilon_{\sigma \rho}$, we have
\begin{align}
% \Boxed{
\delta_{\varepsilon} \vp (x)  \equiv \{\vp (x) , \varepsilon_{\rho \sigma} J^{\rho \sigma}  \}
= \varepsilon_{\rho \sigma} ( x^{\rho} \pa^{\sigma} - x^{\sigma} \pa^{\rho} ) \vp(x)
%}
\, .
\label{eq:PgenLortransf}
\end{align}
In quantum field theory,
the variables $\vp (x)$ and the observables
$P^\m, J^{\rho \sigma} $
become operators,
the Poisson bracket being replaced by $1/\ri \hbar$ times the commutator of operators.

The result~\eqref{eq:PgenTransl} also holds for free spinor fields.
This result states that an infinitesimal, global, Lagrangian symmetry transformation
of fields like
$\delta ^{\textrm{L}}_a \phi \equiv a^\m \pa_\m \phi$ coincides with the infinitesimal Hamiltonian (canonical)
symmetry transformation $\delta ^{\textrm{H}}_a \phi (x) \equiv \{\phi (x) , a^\m P_\m \}$
which is given by the Poisson bracket
 of fields $\phi$ with the
Lagrangian Noether
charges $P_\m$ (expressed in terms of phase space variables $\phi, \pi$).
This fact has actually been proven quite generally in classical
mechanics (see section 7.12.3 of reference~\cite{DeriglazovBook2017}) for the case of non singular Lagrangians
$L(q^i , \dot q ^i)$, i.e. for the case where
$\textrm{det} \, \big( \frac{\pa^2 L}{\pa  \dot q ^i \pa  \dot q ^j} \big) \neq 0$.
The line of arguments of the latter proof should also carry over to continuous systems,
i.e. to non singular Lagrangian field theories as considered above.
However, gauge field theories represent singular dynamical systems. More precisely,
for pure gauge theories, the gauge invariance leads to a so-called \emph{constraint:}
E.g. for the free Maxwell field $(A^\m )$, we have
$\pi_0 \equiv {\pa \cal L}/{\pa \dot{A}^0} = F^{00} =0$
where $F^{\m \n} \equiv \pa^\m A^\n - \pa ^\n A^\m$ are the components of the Faraday tensor.
The space-time translations are then no longer generated by the canonical Noether charge $P^\m$ and
by the canonical Poisson brackets as in equation~\eqref{eq:PgenTransl}:
In a general gauge, the canonical expression for $P^\m$ has to be extended following Dirac's
treatment of constrained Hamiltonian systems so as to \emph{construct a
``kinematical energy-momentum vector'' $P^\m _{\txt{kin}}$ for gauge fields which generates space-time translations of fields.}
If the gauge freedom is completely fixed (e.g. by choosing the radiation gauge or the axial gauge),
then the Poisson brackets have to be
 replaced by the so-called Dirac brackets~\cite{Hanson:1976cn}.
A similar conclusion holds for the Lorentz transformations generated by the components
of the angular momentum: these components are important for instance for the determination of the
 spin of the nucleon,
the latter being made up of the angular momenta of its constituents
(quarks and gluons)~\cite{Leader:2013jra, Wakamatsu:2014zza, Leader:2015vwa}.

To conclude, we note that relations~\eqref{eq:def-H}--\eqref{eq:PgenLortransf}
represent the conventional formulation of Hamiltonian dynamics which is based on Poisson brackets that
are defined at equal times.
This formulation is also referred to as the \emph{instant form} formulation of the classical dynamical system.
As noted by Dirac in 1949~\cite{Dirac:1949cp} (see also~\cite{Hanson:1976cn} and references therein), one may
replace the  hyperplanes
  ``$x^0 =$ constant'' of Minkowski space $\br ^n$ by a family of hypersurfaces defined by a condition
of the form
\begin{align}
F(x) = \tau = \mbox{constant} \, ,
\label{eq:DefTime}
\end{align}
where $F$ denotes a suitably chosen function. For instance, for $F(x) = x^0$, one recovers the conventional
constant time hypersurfaces, and for
\[
F(x) \equiv \frac{1}{\sqrt{2}} \, (x^0 + x^{n-1}) \equiv x^+
\, ,
\]
one obtains the so-called \emph{null-plane} or \emph{light-front formulation} which has received a lot of attention
in the context of two-dimensional conformal field theory, of string theories as well as for gauge field theories in general
dimension, e.g. see references~\cite{\heinzlbakker}.
In the formulation of field theory based on~\eqref{eq:DefTime},
the fields are considered to be functions of ``time'' $\tau$ and of $n-1$ ``spatial'' coordinates $\vec{\sigma}$
which are chosen in such a way that $(\tau , \vec{\sigma} \,)$ parametrizes   Minkowski space.

Following  R.~E.~Peierls~\cite{Peierls:1952cb}, one may also consider
the so-called \emph{Peierls bracket}~\cite{Peierls:1952cb, DeWitt:2003pm}
which represents a Poisson bracket of fields
%$F(t, \vec x \, ), \,  G(t', \vec x ^{\prime} \, )$
at different times. We will come back to this bracket in our concluding remarks.

%%%%%%%%%%%%%%%%%%%%%%%%%%%%%%%%%%%%%%%%%%%%%%%%%%%%%%%%%%%%%%%%%%%%%%%%%%%%%%
%\section{EMT within the Hamiltonian formulation of field theory}\label{sec:HamFormEMT}

%%%%%%%%%%%%%%%%%%%%%%%%%%%%%%%%%%%%%%%%%%%%%%%%%%%%%%%%%%%%%%%%%%%%%%%%%%%
\section{Scalar and Dirac fields}
\label{sec:freefields}

The canonical EMT (energy-momentum tensor) $T_{\txt{can}}^{\mu \nu} [\phi]$
for a \emph{free real massive scalar field} $\phi$ in $\br ^n$,
whose dynamics is described by the Lagrangian density
${\cal L} \equiv \frac{1}{2} \, \left[ (\pa^{\m} \phi) (\pa_{\m} \phi) - m^2 \, \phi^2 \right]$,
%, as given by
%\begin{align}
%T_{\txt{can}}^{\mu \nu} [\phi] & = (\pa^\m \phi) (\pa^\n \phi) - \frac{1}{2} \, \eta^{\m \n}
%\left[  (\pa^{\rho} \phi) (\pa_{\rho} \phi) - m^2 \, \phi^2
% \right]
%\,,
%\end{align}
yields the conserved
energy-momentum vector $P^\nu \equiv \int d^{n-1}x \; T^{0 \nu}_{\txt{can}}$ with
\begin{align}
\label{eq:Pfreescalar}
P^0 = \dfrac{1}{2} \int d^{n-1}x \; \left[ \pi^2 + (\vec{\nabla} \phi)^2 + m^2 \phi^2 \right]  = H
\, , \qquad
\vec P = - \int d^{n-1}\! x \ \pi \, \vec{\nabla} \phi
\, .
\end{align}
From these expressions and definition~\eqref{eq:defpoisson} of the canonical Poisson bracket,
one readily infers that \eqref{eq:PgenTransl} holds for $\vp = \phi$
and $\vp = \pi$ (while taking into account the equation of motion $(\Box +m^2)\phi =0$).

For the \emph{free Dirac field} described by the Lagrangian
\begin{align}
 {\cal L}_{\txt{real}} (\psi ) & \equiv  {\ri} \,
\bar{\psi}  \gamma^{\mu } \!\stackrel{\leftrightarrow}{\pa_\mu} \!\!  \psi -  m \bar{\psi}  \psi
\equiv \frac{\ri}{2}
 \left[ \bar{\psi}  \gamma^{\mu } \pa_\mu \psi - (\pa_\mu \bar{\psi})  \gamma^{\mu }  \psi \right] -  m \bar{\psi}  \psi
\,,
\nn
\\
\txt{or by} \qquad
{\cal L} (\psi ) & \equiv  \bar{\psi} ({\ri}
  \gamma^{\mu } {\pa_\mu}  -  m  )  \psi
\,,
\end{align}
we have
\begin{align}
P^0 = \int d^{n-1}x \; \bar{\psi} ( -\ri \gamma^k \pa_k +m  ) \psi = H
\, , \qquad
\vec P = \int d^{n-1}\! x \ \psi^{\dagger} ( - \ri \vec{\nabla} \psi )
\, ,
\label{eq:PDirac}
\end{align}
and relations~\eqref{eq:PgenTransl} hold for $\vp = \psi_{\alpha}$ and
$\vp = \psi_{\alpha}^* = \ri \pi_{\alpha}^*$.
Since the latter equation represents a relation between phase space variables, it represents strictly speaking
a constraint equation: $\psi_{\alpha}^* - \ri \pi_{\alpha}^* =0$.
However, these constraints for the Dirac field are \emph{second class} and the replacement
of the Poisson bracket by the Dirac bracket~\cite{Das:2008}  yields results of the same form as those obtained by ignoring this subtlety.

We note that relations~\eqref{eq:Pfreescalar} and~\eqref{eq:PDirac}
also imply that the charges $P^\m$
Poisson-commute, i.e. we have an Abelian algebra of charges:
\begin{align}
 \Boxed{
 \{ P^\m, P^\n \} = 0
}
\qquad \mbox{for} \ \; \m, \n \in \{ 0, 1, \dots, n-1 \}
\, .
\label{eq:AbelianAlg}
\end{align}
A consistency check for these results consists in combining~\eqref{eq:PgenTransl}
and~\eqref{eq:Pfreescalar} or~\eqref{eq:PDirac} to verify the Jacobi identity
$0= \{ \vp ,  \{ P^\m, P^\n \} \} + $ cyclic permutations of factors.

To summarize, both for free scalar fields and free Dirac fields
the components $P^\m$ of the canonical energy-momentum vector
generate space-time translations by means of the Poisson brackets, cf. equation~\eqref{eq:PgenTransl}.
This result generalizes to the case where one has a
multiplet of free scalar or free Dirac fields which is invariant
under \emph{global} (rigid) gauge transformations.
Even more generally, one can consider a globally gauge invariant self-interaction of matter fields,
e.g. include a self-interaction potential $V(\phi^{\dagger} \phi)$ for a multiplet of scalar fields
(like the Higgs field) or an invariant Yukawa-type coupling between scalar and spinor fields.
However, if gauge fields are involved, things become more subtle as we will discuss  in the next section.

%%%%%%%%%%%%%%%%%%%%%%%%%%%%%%%%%%%%%%%%%%%%%%%%%%%%%%%%%%%%%%%%%%%%%%%%%%%%%
%%%%%%%%%%%%%%%%%%%%%%%%%%%%%%%%%%%%%%%%%%%%%%%%%%%%%%%%%%%%%%%%%%%%%%%%%%%
\section{Lagrangian formulation of pure gauge theories}\label{sec:LagrForm}

%%%%%%%%%%%%%%%%%%%%%%%%%%%%%%%%%%%%%%%%%%%%%%%%%%%%%%%%%%%%%%%%%%%%%%%%%%%%%
\subsection{General set-up}\label{sec:GenSetUp}

In the sequel we are interested in pure Abelian gauge theory (free Maxwell theory) and in pure non-Abelian gauge theory
(pure YM theory). For concreteness we will consider the \emph{four dimensional} case and in order to
avoid redundancies in the presentation, we will present the generalities for the case of a general symmetry group $G$,
Maxwell's theory corresponding to the particular case $G = U(1)$.

More precisely,
as symmetry group we consider a compact, semi-simple matrix  Lie group $G$
of dimension $n_G$ and we denote the associated Lie algebra by $\lie$. The
gauge potential is given by a  $\lie$-valued vector field ${A} _{\mu} (x) \equiv
A_{\mu} ^a (x) T^a $.
Here, $(A_\m ^a )_{\mu \in \{ 0,1,2,3 \}}$
is a real-valued
vector field in four space-time dimensions
 for each value of the internal index $a \in \{ 1, \dots , n_G \}$ and $\{ T^a \}_{a \in \{ 1, \dots , n_G \}}$
 is a basis of the Lie algebra $\lie$.
 We have
\begin{equation}
[  T^a , T^b ]  = \ri f^{abc}  T^c
\, ,
\label{eq:LArelations}
\end{equation}
where the real structure constants $f^{abc}$ can be chosen to be totally antisymmetric in the indices
for semi-simple Lie algebras, e.g. $su( N )$.
Under an infinitesimal gauge variation parametrized by a
 $\lie$-valued function $x \mapsto \omega (x) \equiv \omega ^a(x) T_a$,
 the gauge potential transforms with the covariant derivative of $\omega$:
\begin{equation}
\label{eq:CDom}
\delta A_\m = D_\m \omega \equiv \pa_\m \omega  + \ri \mq [ A_\m , \omega ]
\, .
\end{equation}
Here, the coupling constant $\mq$ represents the
``non-Abelian'' or ``YM'' charge.

The  $\lie$-valued field strength tensor associated to the gauge potential $A_\m$ reads
${F} _{\mu \nu} \equiv {F} _{\mu \nu}^a T_a$ with
${F} _{\mu \nu} \equiv \pa_{\mu} {A} _{\nu} - \pa_{\nu}
{A} _{\mu} + \ri \mq \, [ {A} _{\mu}  ,  {A} _{\nu} ]$.
As usual, the components of this tensor will be denoted by
\[
F^{i0} = E_{i} \, , \qquad F^{ij} = - \varepsilon^{ijk} B_{k}
\qquad (\, \mbox{with the normalization} \ \, \varepsilon^{123} =1 \, )
\, .
\]
In the non-Abelian case, the vector fields $\vec E \equiv (E_i)_{i=1,2,3}$ and $\vec B \equiv (B_i)_{i=1,2,3}$
represent the chromo-electric and chromo-magnetic fields\footnote{The notation
$E_i$ for $E_{x^i}$ is convenient, but it should be kept in mind
in this context that $i$ is not a covariant (Lorentz) index since $\vec E$ is not
the spatial part of a four-vector (and similarly for $B_i$).}, respectively.

For \emph{free Maxwell theory,} the internal index takes a single value
$a=1$ and the totally antisymmetric structure constants $f^{abc}$ in~\eqref{eq:LArelations} vanish,
as does the commutator term in the field strength ${F} _{\mu \nu}$ and in the covariant derivative~\eqref{eq:CDom}.
In this case, the vectors $\vec E$ and $\vec B$
represent the electric and magnetic fields, respectively and
there is presently no self-interaction of gauge potentials in the action~\eqref{eq:invact} below.

%%%%%%%%%%%%%%%%%%%%%%%%%%%%%%%%%%%%%%%%%%%%%%%%%%%%%%%%%%%%%%%%%%%%%%%%%%%%%
\subsection{Lagrangian formulation}\label{sec:LagFormulGT}

%%%%%%%%%%%%%%%%%%%%%%%%%%%%%%%%%%%%%%%%%%%%%%%%%%%%%%%%%%%%%%%%%%%%%%%%%%%%%
\paragraph{Dynamics:}
The dynamics of pure gauge theory is described by the classical action
\begin{equation}
S [ A]  \equiv   -  \frac{1}{4} \, \int d^4 x \ \Tr
\, ( {F} ^{\mu \nu} {F} _{\mu \nu} )
 = -
 \frac{1}{4}  \int d^4 x
 \,{F} ^{a \mu \nu}
\, {F}^a _{\mu \nu}
=  \frac{1}{2} \, \int d^4 x \ \Tr
\, ( \vec E ^{\, 2} - \vec B ^{\, 2} )
 \, ,
 \label{eq:invact}
\end{equation}
where we absorbed the so-called \emph{index} of the considered Lie algebra representation
into the definition of the trace.
The functional~\eqref{eq:invact} is gauge invariant and its variation yields the YM field equation
$0 = D_\n F^{\n \m} = \pa_\n F^{\n \m} + \ri q [A_\n , F^{\n \m} ]$.
In terms of the chromo-electric and -magnetic fields, the latter equations read
\[
D_i E_i =0 \, , \qquad
\varepsilon^{ijk} D_j B_{k} = D_0 E_i
\, .
\]
In the Abelian case, these equations represent Maxwell's equations
 $ \Div \, \vec E=0$ and $\curl \, \vec B = \pa_t {\vec E}$.

%%%%%%%%%%%%%%%%%%%%%%%%%%%%%%%%%%%%%%%%%%%%%%%%%%%%%%%%%%%%%%%%%%%%%%%%%%%%%%%%%%%%%%%%%%%%%%%%%%
\paragraph{Translational invariance:}
By virtue of Noether's first theorem,
the invariance of the action functional $S[A] \equiv \int d^4 x \, {\cal L}$
given by~\eqref{eq:invact}
under \emph{space-time translations}
\begin{align}
\label{eq:TransTransA}
\delta x^\m = a^\m \, , \qquad
\delta A^\m = - a^\n \pa_\n A^\m
\,,
\end{align}
implies the local conservation law $\pa_\m  T_{\txt{can}}^{\m \nu}  =0$
for the solutions of the (Lagrangian) equations of motion $D_\n F^{\n \m} =0$. Here,
\begin{align}
T_{\txt{can}}^{\mu \nu}  \equiv
\frac{\pa {\cal L}}{\pa (\pa_{\mu} A_\rho ^a )} \; \pa^\n   A_\rho ^a -\eta^{\m \n} {\cal L} =
  \Tr \, ( -  F^{\mu \rho}  \, \pa^{\nu}  A_\rho   + \frac{1}{4} \, \eta^{\mu \nu} F^{\rho \sigma}  F_{\rho \sigma} )
\, ,
\label{eq:canEMTYM}
\end{align}
represents the \emph{canonical EMT}.
By using  the  equations of motion, this tensor may be decomposed
into a gauge invariant part $T_{\txt{inv}} ^{\m \nu}$
(generally referred to as the \emph{improved EMT} of the gauge
 field~\cite{Blaschke:2016ohs})
and a \emph{superpotential term} (i.e. a total derivative $ \pa_{\rho}  \chi ^{\m \rho \n}$ with $\chi ^{\m \rho \n} = - \chi ^{\rho \m \n}$):
 \begin{align}
T_{\txt{can}} ^{\m \nu} & =
T_{\txt{inv}} ^{\m \nu}
+ \pa_{\rho}  \Tr \, ( - F^{\m \rho} A^\n )
\, ,  \qquad \mbox{with} \quad
 \Boxed{
 T_{\txt{inv}} ^{\m \nu}  =
  \Tr \,  ( F^{\m \rho} {F_{\rho}} ^\n + \frac{1}{4} \, \eta ^{\m \n} F^{\rho \sigma} F_{\rho \sigma} )
  }
\, .
\label{eq:impEMTYM}
\end{align}
Accordingly, the EMT $T_{\txt{inv}}^{\m \nu} $ is also locally conserved
for the  solutions of the equations of motion and, for these solutions, it
yields the same conserved charges as $T_{\txt{can}}^{\m \nu}$
(for fields which vanish sufficiently fast at spatial infinity).
More precisely, for the  solutions of the equations of motion, the \emph{canonical Noether charges} $ P^\n$
and the  \emph{gauge invariant charges}  $ P^\n _{\txt{inv}}$
are related by
\begin{align}
P^\n  & \equiv \int d^3 x \, T^{0\n} _{\textrm{can}}
= P^\n _{\txt{inv}} + \int d^3 x \, \pa_j   \Tr \,  ( - F^{0 j} A^\n    )
\, , \qquad  \mbox{with} \quad
P^\n _{\txt{inv}}  \equiv \int d^3 x \, T^{0\n} _{\textrm{inv}}
\, .
\end{align}
The explicit expressions have the form
\begin{align}
\Boxed{
P^0  =
\int d^3 x \;  \Tr \,  \big[  \frac{1}{2} \, (\vec E ^{\, 2} + \vec B^{\, 2} ) - A^0 \, (D_i E_i)  \big]
= H
}
\, , \qquad
\Boxed{
\vec{P} =  \int d^3 x \; \Tr \,  (  E_{i} \vec{\nabla} A^i )
}
\label{eq:canonP}
\, ,
\end{align}
and
\begin{align}
\Boxed{
P^0_{\txt{inv}} =\frac{1}{2}  \int d^3 x \;  \Tr \,  (   \vec E ^{\, 2} + \vec B^{\, 2} )  = H_{\txt{inv}}
}
\, , \qquad
\Boxed{
\vec{P}_{\txt{inv}} \equiv  \int d^3 x \;  \Tr \,  (  \vec E \times \vec B \,  )
}
\label{eq:gaugeinvP}
\, .
\end{align}
Here, $\frac{1}{2} \,   \Tr \, (\vec E ^{\, 2} + \vec B^{\, 2})$ represents
the total energy density of the fields and $\Tr \, (\vec E \times \vec B \, )$ the associated Poynting vector.
We note that the result~\eqref{eq:impEMTYM} as well as the conserved currents associated to the invariance 
under Lorentz transformations in $\br^n$ and under scale transformations in $\br^4$ 
are derived in a concise and unified manner in appendix \ref{sec:Derivation}.

%%%%%%%%%%%%%%%%%%%%%%%%%%%%%%%%%%%%%%%%%%%%%%%%%%%%%%%%%%%%%%%%%%%%%%%%%%%%%%%%%%%%%%%%%%%%%%%%%%
\paragraph{Lorentz invariance:}
The invariance
 of the action functional $S[A] \equiv \int d^4 x \, {\cal L}$ under \emph{Lorentz transformations}
\begin{align}
\delta x^\m = 2 {\varepsilon^{\m}}_{ \n} x^\n\, , \qquad
\delta A^\m =
\varepsilon_{\rho \sigma} \big[ (x^\rho \pa^\sigma - x^\sigma \pa ^\rho ) A^\m +
\eta^{\m \rho} A^\sigma - \eta^{\m \sigma} A^\rho \big]
\qquad \mbox{with}  \ \; \varepsilon^{\m \n} = -  \varepsilon^{\n \m}
\label{eq:LtransforA}
\end{align}
implies the local conservation law $\pa_\m  M_{\txt{can}}^{\m \rho \sigma}  =0$
for the  solutions of the equations of motion.
Here,  $ M_{\txt{can}}^{\m \rho \sigma}$ denotes the
\emph{canonical angular momentum tensor} of the gauge field:
\begin{align}
\label{eq:CanAMTgf}
M_{\txt{can}}^{\m \rho \sigma} & \equiv
x^\rho  T_{\txt{can}}^{\m \sigma} - x^\sigma T_{\txt{can}}^{\m \rho}
+   \Tr \, ( - F^{\m \rho} A^\sigma +  F^{\m \sigma } A^\rho )
\, .
\end{align}
For the  solutions of the equations of motion, the latter can be decomposed into
 a gauge invariant part and a
superpotential term:
\begin{align}
M_{\txt{can}}^{\m \rho \sigma} & =
 M_{\txt{inv}}^{\m \rho \sigma}
+   \pa_{\n} \Tr \, \big[ \! - F^{\m \nu}  (x^\rho A^\sigma - x^\sigma A^\rho ) \big]
\, , \qquad \mbox{with} \quad
\Boxed{
 M_{\txt{inv}}^{\m \rho \sigma}  =
x^\rho  T_{\txt{inv}}^{\m \sigma} - x^\sigma T_{\txt{inv}}^{\m \rho}
}
\, .
\label{eq:MaxAMT}
\end{align}
For the \emph{canonical conserved charges}
$J^{\rho \sigma} \equiv  \int d^3 x \, M_{\txt{can}}^{0\rho \sigma}$,
it is convenient to introduce the following notation for the purely spatial parts:
$J ^i \equiv \frac{1}{2} \, \varepsilon^{ijk}J ^{jk}$ and
$\vec J \equiv (J ^i)_{i=1,2,3}$.
We have
\begin{align}
\label{eq:CanNClorentz}
J^{ij} & =  \int d^3 x \; \Tr \, \big[  E_{k} (x^i \pa^j - x^j \pa^i) A_k + E_{i} A^j -  E_{j} A^i \big]
\,,
\end{align}
\begin{align}
\mbox{or} \quad
\Boxed{
\vec J  = \vec L + \vec S
}
\qquad \mbox{with} \ \;
\Boxed{
\vec L = \int d^3 x \; \Tr \, \big[ E_k (\vec x \times \vec{\nabla} ) A^k \big]
}
\, , \quad
\Boxed{
\vec S = \int d^3 x \; \Tr \, \big( \vec E \times \vec A \big)
}
\, ,
\label{eq:CanJLS}
 \end{align}
and
\begin{align}
& J^{0i}  =  \int d^3 x \; \Tr \, \big[  E_{k} (x^0 \pa^i - x^i \pa^0) A_k -  E_{i} A^0
+  x^i \, \frac{1}{2} (\vec E ^{\, 2} - \vec B^{\, 2} ) \big]
\,,
\label{eq:JoiLag}
\\
\mbox{or} \quad  &
%\Boxed{
J^{0i}
=  \int d^3 x \; \Tr \,  \big[ x^0 E_{k} \pa^i A_k
 - x^i \, \frac{1}{2} \, (\vec E ^{\, 2} + \vec B^{\, 2} )
 + x^i A^0 \, (D_k E_k)
\, \big]
%}
\, .
\label{eq:JoiHam}
\end{align}
For the solutions of the Lagrangian equations of motion $D_\n F^{\n \m} =0$,
the canonical charges $J^{\rho \sigma}$ coincide with
 \emph{gauge invariant charges}
$J_{\txt{inv}} ^{\rho \sigma} \equiv  \int d^3 x \, M_{\txt{inv}}^{0\rho \sigma}$
which read as follows (in terms of the notation
$\vec J_{\txt{inv}} \equiv (J_{\txt{inv}} ^i)_{i=1,2,3}$
with $J_{\txt{inv}} ^i \equiv \frac{1}{2} \, \varepsilon^{ijk}J_{\txt{inv}} ^{jk}$):
\begin{align}
%\Boxed{
\vec J_ {\txt{inv}}=  \int d^3 x \; \Tr \, \big[ \vec x \times (\vec E \times \vec B) \big]
%}
\, , \qquad
%\Boxed{
 J^{0i}_ {\txt{inv}}=  \int d^3 x  \; \Tr \,  \big[  x^0  (\vec E \times \vec B)_i - x^i
\, \frac{1}{2}   \, (\vec E ^{\, 2} + \vec B^{\, 2}) \big]
%}
\, .
\label{eq:GenLorTr}
\end{align}
Thus, we have $\vec J_ {\txt{inv}} = \vec J = \vec L + \vec S$ for the solutions of the equations of motion, but $\vec L$ and $\vec S$
are not gauge invariant.
As a matter of fact,
it is not possible to decompose the total angular momentum in a gauge invariant manner into orbital and spin parts --- see reference~\cite{Jauch:1976ava}
 for this issue and for the related question
of how to attribute
a gauge invariant, and thereby physical, meaning to the spin of the photon.

%%%%%%%%%%%%%%%%%%%%%%%%%%%%%%%%%%%%%%%%%%%%%%%%%%%%%%%%%%%%%%%%%%%%%%%%%%%%%%%%%%%%%%%%%%%%%%%%%%
\paragraph{Conserved charges associated to Poincar\'e invariance:}
The structure of the generators~\eqref{eq:CanJLS}--\eqref{eq:GenLorTr} can nicely be exhibited
by using the notation
\[
 H \equiv  \int d^3 x \, {\cal H} \, , \quad
H  _{\txt{inv}} \equiv  \int d^3 x \, {\cal H}_ {\txt{inv}} \, , \quad
\vec P \equiv  \int d^3 x \, \vec{{\cal P}} \, , \quad
\vec P _ {\txt{inv}} \equiv  \int d^3 x \, \vec{{\cal P}} _ {\txt{inv}} \, ,
\]
and
\[
\vec K \equiv  \left( {J}^{0i} \right)_{i =1,2,3}  \, , \qquad
\vec K _{\txt{inv}} \equiv  \left( {J}_{\txt{inv}}^{0i} \right)_{i =1,2,3}
\, ,
\]
in terms of which we have
\begin{align}
\Boxed{
\vec L =  \int d^3 x \; \vec x \times \vec{\cal P}
}
\, , \qquad
&
\Boxed{
\vec K = \int d^3 x \,  \big(  t \, \vec{\cal P} - \vec x \, {\cal H}  \big)
}
\,,\\
\Boxed{
\vec J_ {\txt{inv}} =  \int d^3 x \; \vec x \times \vec{\cal P} _{\txt{inv}}
}
\, , \qquad
&
\Boxed{
\vec K _ {\txt{inv}}  = \int d^3 x \,  \big(  t \, \vec{\cal P}_ {\txt{inv}}  - \vec x \, {\cal H}_ {\txt{inv}}   \big)
}
\, .
\label{eq:GenLor}
\end{align}

%%%%%%%%%%%%%%%%%%%%%%%%%%%%%%%%%%%%%%%%%%%%%%%%%%%%%%%%%%%%%%%%%%%%%%%%%%%%%%%%%%%%%%%%%%%%%%%%%%
\paragraph{Conserved charges associated to conformal invariance:}
We note that, in four space-time dimensions,
pure Abelian or non-Abelian gauge field theory is not only invariant under the Poincar\'e group,
but also under the larger \emph{conformal group} (see appendix \ref{sec:Derivation} for the details):
 The generators corresponding to dilatations and special conformal transformations can be treated
along the same lines.

%%%%%%%%%%%%%%%%%%%%%%%%%%%%%%%%%%%%%%%%%%%%%%%%%%%%%%%%%%%%%%%%%%%%%%%%%%%%%%%%%%%%%%%%%%%%%%%%%%
\paragraph{Other conserved charges:}
For completeness, we mention that a free field theory admits an infinite number of conserved current
densities~\cite{Kibble1965, Steudel66, Olver}.
In particular, for free Maxwell theory, one can find gauge invariant conserved quantities differing from those considered above.
The latter include in particular the so-called \emph{zilch} currents introduced by D.~M.~Lipkin~\cite{Lipkin1964}
(see also~\cite{Kibble1965, Gordon1984}),
i.e. the current densities
\[
Z^\m _{\ \n \rho} \equiv \tilde{F} ^{\m \lambda}  \!\stackrel{\leftrightarrow}{\pa_\rho} F_{\lambda \n} +
\tilde{F} _{\n \lambda}  \!\stackrel{\leftrightarrow}{\pa_\rho} F^{\lambda \m} \, ,
\qquad \mbox{where} \ \; \tilde{F} ^{\m \n}  \equiv \frac{1}{2} \, \varepsilon ^{\m \n \rho \sigma} F_{\rho \sigma}
\, .
\]
For  their potential physical relevance (in particular
the relationship with the optical chirality) and for the underlying symmetry transformations, we refer to~\cite{Philbin2013} and references therein.
We also mention that related conserved currents involving spatial non-localities can be found and have been discussed
for Maxwell theory~\cite{Bernabeu:2019dym}. The latter result in particular from the \emph{duality rotations}
parametrized by $\th \in \br$,
\[
\vec E ^{\prime} = \cos \th \, \vec E + \sin \th \, \vec B \, , \qquad
\vec B ^{\prime} = - \sin \th \, \vec E + \cos \th \, \vec B  \, ,
\]
which leave the Maxwell equations (and action) invariant~\cite{Calkin1965, Deser1976, Deser1982, Bunster:2011aw, Cameron2012}.
Yet, it should be stressed that the conservation laws which hold for a free field theory do in general not carry over to the interacting theory~\cite{Kibble1965}.

\section{Hamiltonian formulation of free Maxwell theory}\label{sec:HamFormMaxwell}
%%%%%%%%%%%%%%%%%%%%%%%%%%%%%%%%%%%%%%%%%%%%%%%%%%%%%%

The Hamiltonian formulation being more subtle than the Lagrangian one, we first discuss the Abelian theory in this section
and then point out
the essential modifications which are brought about by the non-Abelian theory in  section~\ref{sec:HamYM}.
 We again consider the four dimensional case while starting from the action functional~\eqref{eq:invact}.

\subsection{Canonical momenta and Hamiltonian}
%%%%%%%%%%%%%%%%%%%%%%%%%%%%%%%%%%%%%%%%%%%%%%%%%%%%%%
The momentum which is canonically conjugate  to $A _{\mu}$ is defined by
 $\pi^{\mu} \equiv \pa {\cal L} / \pa \dot{A} _{\mu} = F^{\mu0}$, hence
 $\vec{\pi}$ coincides with the electric field strength $\vec{E}$:
\begin{align}
 \label{eq:piA}
 \Boxed{
\pi^k = F^{k0} = E_{k}
}
\, .
\end{align}
From $E_k =  F^{k0} = \pa^k A^0 - \pa^0 A^k = ( - \vec{\nabla} A^0 - \pa_t \vec A \, )^k$,
it follows that the relation $\pi^{k} = \pa {\cal L} / \pa \dot{A} _{k}$
can be solved for the derivative $\dot{A}_k$ in terms of $\vec{\pi}$ and
$\vec{\nabla} A^{0}$: $\dot{\vec A}= - \vec{\pi} - \vec{\nabla} A^0$.

 From the antisymmetry of the tensor field $(F^{\mu \nu})$ it follows that
$\pi^0 = F^{00} =0$.
Henceforth we cannot use the explicit expression for $\pi^0$ to
express the time derivative $\dot A ^0$ in terms of the fields
$\pi^\m$
(and, possibly, $A^\m$
and/or the spatial derivatives of $A^\m$)  as required by the standard Hamiltonian formulation\footnote{The
fact that the Lagrangian density
does not depend on $\dot{A}^0$ reflects a degeneracy
(related to gauge invariance) and means that $A^0$ does not really represent a dynamical variable.}.
Thus, the dynamical system under consideration represents a so-called
\emph{constrained Hamiltonian system} \cite{Dirac:1964,Hanson:1976cn,\GitmanProkhorov,Henneaux:1992,Wipf:1993}:
We have the relation
\begin{equation}
\label{primconsem}
0 =
\phi_1 \equiv \pi^0
\, ,
\end{equation}
whose origin can be traced back to the invariance of the Lagrangian under gauge transformations,
e.g. see reference~\cite{Wipf:1993}.
Since the condition~\eqref{primconsem} results directly from the Lagrangian,
it is referred to as a \emph{primary constraint}.
(As a matter of fact, $\pi^0$ depending on the space-time coordinates,
this equation actually represents an infinity
of constraints, one for each $\vec x \in \br ^3$.)

The \emph{canonical Hamiltonian function}
$H\equiv P^0 \equiv \int d^3 x \, T^{00}_{\textrm{can}}$ reads
\begin{equation}
\label{canhamem}
H \equiv \int d^3 x \, {\cal H} \equiv
\int d^3 x \, ( \dot{A} _{\mu} \pi^{\mu} - {\cal  L}   )
= \int d^3 x \, \left[
 \frac12 \, \pi^i \pi^i
+ \frac{1}{4} \, {F} _{ij} {F} ^{ij}
- A_0 (\pa_i \pi^i ) \right]
\, ,
\end{equation}
where the form of the last term results from an integration by parts, assuming as usual that fields vanish at infinity.
After substituting
 $F^{ij} = - \varepsilon^{ijk} B_{k}$
%(with $\vec B = \curl \, \vec A$)
and $\pi^i = E_{i}$, we recognize the Abelian special case of
 expression~\eqref{eq:canonP} for the canonical Hamiltonian of YM theory.

%%%%%%%%%%%%%%%%%%%%%%%%%%%%%%%%%%%%%%%%%%%%%%%%%%%%%%%%%%%%%%%%%%%%%%%%%%%%%%%%%%%%%%%%%
\subsection{Canonical Poisson brackets}
%%%%%%%%%%%%%%%%%%%%%%%%%%%%%%%%%%%%%%%%%%%%%%%%%%
The fundamental Poisson-commutator  for the canonically conjugate pair $(A, \pi )$, which holds for any fixed time $t$,
reads
\begin{equation}
\label{fundpoissbra}
\{ A_{\mu} (t, \vec x \, ) , \pi^{\nu} (t , \vec y \, ) \} = \delta_{\mu}^{\nu} \, \delta (\vec x - \vec y \, )
\, .
\end{equation}
In particular we have
\begin{align}
\label{eq:PBAE}
\{ A_{i} (t, \vec x \, ) , E_{k} (t , \vec y \, ) \} & = \delta_{i}^{k} \, \delta (\vec x - \vec y \, )
\,,\\
\mbox{whence} \quad \{ E_{i} (t, \vec x \, ) , B_{j} (t , \vec y \, ) \} & = \varepsilon_{ijk} \pa_k^{(x)} \, \delta (\vec x - \vec y \, )
\, ,
\nn
\end{align}
(where the last relation follows from the first one by virtue of $\vec B = \curl \, \vec A$),
all other brackets between $A^\m, \pi^\n$ and $B_{k}$ vanishing.

A few remarks concerning these brackets are in order.
First, we note that relations~\eqref{eq:PBAE} are sufficient for evaluating brackets  between functionals which only depend on $\vec A, \vec E$ and $\vec B$.
For instance, one finds that the brackets between the components of the spin angular momentum $\vec S$, as defined in equation~\eqref{eq:CanJLS},
satisfy the Lie algebra of infinitesimal rotations,
\begin{align}
\label{eq:BracketSS}
\{ S^i  , S^j  \} & = \varepsilon^{ijk} \, S^k
\, ,
\end{align}
and similarly for the components $L^i$ of $\vec L$ and  $J^i$ of $\vec J$.

Second, we stress that for the proper evaluation of Poisson brackets like $\{ A_\m (t , \vec x \, ) , J^{0i} \}$,
 the functional $J^{0i}$ (as given by expression~\eqref{eq:JoiLag} or~\eqref{eq:JoiHam})
 has to be expressed in terms of canonical variables, i.e.
$\pa_0 A^k$ has to be rewritten in terms of $\pi^k = E_k$ and $\pa_k A^0$.

Finally, we note that the basic relation~\eqref{fundpoissbra} is not compatible with the constraint $\pi^0 =0$. In fact,
an {important point} in this context is that
%all dynamical variables are allowed to vary with time and that
\emph{the constraint equations must not be substituted into
the Poisson brackets:}
they can only be imposed after computing the Poisson brackets and then amount to projecting
the result onto the constraint surface.
Indeed, generalizing earlier work of P.~Bergmann and his collaborators, Dirac devised a general approach
to handle constrained Hamiltonian systems \cite{Dirac:1964,Hanson:1976cn,\GitmanProkhorov,Henneaux:1992,Wipf:1993}
which we will also follow below.

%%%%%%%%%%%%%%%%%%%%%%%%%%%%%%%%%%%%%%%%%%%%%%%%%%%%%%%%%%%%%%%%%%%%%%%%%%%%%%%%%%%%%%%%%
\subsection{Dirac's method and extended Hamiltonian}\label{sec:DirMeth}
%%%%%%%%%%%%%%%%%%%%%%%%%%%%%%%%%%%%%%%%%%%%%%%%%%
For a function $F$ on the phase space $\{ (A^\m , \pi_\m )\}$ which vanishes on the primary constraint surface
defined by relation~\eqref{primconsem},  Dirac introduced the notation $F\approx 0$ (``$F$ vanishes weakly''),
hence we can also write $0 \approx  \phi_1 \equiv \pi^0$.
The first step of Dirac's procedure consists of
including  the primary constraint into the Hamiltonian
by means of an undetermined Lagrange multiplier field $\l^1(x)$ (which comes without associated momentum):
\begin{equation}
H_p \equiv  H + \int d^3 x \, \lambda^1 (x) \phi_1 (x)
\, .
\end{equation}
The second step is to impose that the primary constraint is preserved by the time evolution defined
by the \emph{primary Hamiltonian} $H_p$, i.e. a  stability condition for this constraint:
\[
0 \stackrel{!}{\approx} \dot{\phi} _1 \approx \{\phi_1  , H_p \}
= \pa_i \pi^i
\, .
\]
Obviously this condition gives rise to the so-called \emph{secondary constraint}
\begin{equation}
\label{gauss}
0 \approx \phi_2 \equiv \pa_i \pi^i = \Div \, \vec E
\, .
\end{equation}
%where $\vec E$ denotes the electric field strength.
For the free field theory under consideration, the relation $ \Div \, \vec E =0$
represents the Maxwell equation describing the Gauss law in vacuum and (\ref{gauss})
is therefore referred to as \emph{Gauss law constraint}.
Its time evolution does not give rise to further constraints
since
$\dot{\phi} _2 \approx \{\phi_2  , H_p \} = - \pa_i \pa_j F^{ij} =0$.
Since $\{ \phi_1 , \phi_2 \} = 0$, the constraints $\phi_1, \phi_2$ are
referred to as \emph{first class constraints (FCC's).}

The Hamiltonian equations of motion determined by $H_p$ are equivalent to the Lagrangian equations of motion
of the dynamical system.
For the study of symmetries in the Hamiltonian framework,
Dirac considered a generalization of the Lagrangian formalism given by
the so-called \emph{extended Hamiltonian}
 \begin{equation}
 \label{exthamem}
H_E \equiv H + \sum_{j=1}^2 \int d^3 x \,  \lambda^j \phi_j
=
\int d^3 x \, \left[  \frac{1}{2} \, (\vec E ^{\, 2} + \vec B^{\, 2})
+ \lambda^1 \pi^0 + (\lambda^2 -A_0) \, \Div \, \vec E
\right]
\, ,
\end{equation}
which is defined on the \emph{extended phase space} $\{ (A_\m, \pi^\m , \lambda ^j )\}$.
The Hamiltonian~\eqref{exthamem} involves a linear combination of all of the FCC's $\phi_j$ (the primary and secondary
FCC's being treated on an equal footing) involving  Lagrange multipliers  $\lambda^1$ and $\l^2$.
Geometrically speaking,
the extended Hamiltonian generates all possible flows of the system under consideration,
namely the time evolution determined by the canonical Hamiltonian $H$ as well as the flow generated by the
FCC's (i.e. vanishing conserved quantities)~\cite{Gambini:2011}.

The \emph{extended Hamiltonian equations} are given by
 $\phi_1 = 0 = \phi_2$ and
 $ \dot F = \{ F, H_E \}$ for
 $F \in \{ A^0, \dots , A^3, \pi_0 , \dots, \pi_3 \}$:
for the extended Hamiltonian~\eqref{exthamem}, we thus obtain
\begin{align}
\label{exteom}
\dot{A} ^0 &= \lambda ^1 \, ,  &
\dot{\pi} _0 &= \Div \, \vec E\,, \nn\\
\dot{\vec A}  &= - \vec{\pi} - \grad\, A^0 + \grad\, \lambda ^2
\, ,  &
\dot{\vec{\pi}} &= \curl \, \vec B  \, .
\end{align}
If we substitute the expressions
 $\pi_0 =0$ and $\vec{\pi} = \vec E$ into these relations,
then we recognize the equations for $\dot{\pi} _\m$ as Maxwell's free field equations $\pa_\m F^{\m \n}=0$.
The equation of motion $\dot{A} ^0 = \lambda ^1$ determines $\l^1$ in terms of $\dot{A} ^0$:
we will come back to this relation below.
Here we only note that the last term in~\eqref{exthamem} shows that the field $A^0$ amounts to a Lagrange multiplier
combining with the multiplier $\lambda^2$ which also represents an arbitrary function.

%%%%%%%%%%%%%%%%%%%%%%%%%%%%%%%%%%%%%%%%%%%%%%%%%%%%%%%%%%%%%%%%%%%%%%%%%%%%%%%%%%%%%%%%%
\subsection{Hamiltonian gauge symmetries}
%%%%%%%%%%%%%%%%%%%%%%%%%%%%%%%%%%%%%%%%%%%%%%%%%%
By definition, the \emph{Hamiltonian gauge symmetries} (which are parametrized at the infinitesimal level by ${\varepsilon}^{j}(x)$)
\emph{are generated by the FCC's $\phi_1$ and $\phi_2$:}
for a functional $F$ of $(A, \pi )$, we have
\begin{equation}
 \label{hamgaugesymm}
\delta _{(\varepsilon^j )} F = \{ F , G _{(\varepsilon^j )} \} \, ,
\qquad \txt{with} \ \;
G _{(\varepsilon^j )} \equiv \sum_{j=1}^2 \int d^3 x \,  \varepsilon^j (x)  \phi_j (x) \, .
\end{equation}
By virtue of~\eqref{fundpoissbra}, this relation readily leads to
 $\delta _{(\varepsilon^j )} \pi^{\mu} =0$ and
 \begin{equation}
 \label{varexem}
\delta _{(\varepsilon^j )} A_0 = \varepsilon^1 \, , \qquad
\delta _{(\varepsilon^j )} A_i = -\pa_i \varepsilon^2
\quad \ \txt{for} \ \; i \in \{ 1,2 ,3 \}
\, .
\end{equation}
One can show on general
grounds~\cite{Henneaux:1992} (and easily check for the case at hand)
that the extended equations of motion~\eqref{exteom} are invariant under these
gauge transformations if \emph{the Lagrange multiplier fields $\lambda^j$ transform as}
\begin{equation}
 \label{hamgaugesymmetr}
\delta _{(\varepsilon^j )} \lambda ^1  =  \dot{\varepsilon} ^1  \, ,
\qquad \delta _{(\varepsilon^j )} \lambda ^2  = \dot{\varepsilon} ^2  + \varepsilon^1
 \, .
\end{equation}
For the present dynamical system (for which the Hamiltonian (\ref{canhamem}) is quadratic in the momenta,
and the constraints \eqref{primconsem}, \eqref{gauss}
are linear in the momenta), the Hamiltonian gauge symmetries generated by the FCC's yield
\emph{the Lagrangian symmetries}~\cite{Wipf:1993},
i.e. the infinitesimal gauge transformations $ \delta A_{\mu} = \pa_{\mu} \epsilon$.
Indeed, according to~\eqref{varexem},
the latter transformation laws \emph{are recovered for the choice of parameters
 $\varepsilon^1 \equiv \pa_0 \epsilon, \,  \varepsilon^2 \equiv -\epsilon$,}
 which implies (by virtue of~\eqref{hamgaugesymmetr})
 $\delta _{(\varepsilon^j )} \lambda^2 =0$
 and $\delta _{(\varepsilon^j )} \lambda^1 =\ddot{\epsilon}$.
 This choice of parameters amounts to imposing the following \emph{generalized gauge condition,} i.e.
 gauge fixing condition for the Hamiltonian gauge symmetries~\eqref{varexem}, \eqref{hamgaugesymmetr}:
 \begin{equation}
 \label{eq:gfcondition}
  \mbox{gauge condition} \qquad \lambda^2 =0
  \, .
 \end{equation}
 In this case, the extended Hamiltonian reduces to the primary Hamiltonian
 (which is equivalent to the Lagrangian formulation)
 and the \emph{gauge generator} $G _{(\varepsilon^j )} $ \emph{reduces to the one of
 Lagrangian gauge transformations} which reads
 \begin{align}
 \label{eq:gaugegener}
 G_{\epsilon}   = \int d^3 x \, \left[  (\pa_0 \epsilon) \pi^0 - \epsilon (\pa_i \pi^i) \right]
 = \int d^3 x \, \pi^{\mu}  \pa_{\mu} \epsilon
 \, .
 \end{align}
Thus, with~\eqref{eq:gfcondition} we have
 \begin{align}
 \label{eq:infgaugetrafoHam}
 \delta _{\epsilon} A_\m
 \equiv \{ A_\m , G_{\epsilon} \} =  \pa_\m \epsilon
 \, , \qquad
 \delta _{\epsilon} \pi^\m
 \equiv \{ \pi^\m , G_{\epsilon} \} =  0
 \, .
 \end{align}

 \paragraph{Conclusion: }
This means that \emph{the Lagrangian gauge transformations are the residual Hamiltonian gauge transformations in the
generalized gauge in which the Lagrange multiplier associated to the secondary constraint is put to zero}.
The Lagrange multiplier $\lambda^1$ associated with the primary FCC $\pi ^0 \approx 0$
still remains undetermined. It can be determined by imposing a consistent generalized gauge condition while leaving
the Lagrangian gauge freedom unfixed (see next subsection)
or it can be determined as a consequence of a complete gauge fixing
of the Lagrangian gauge freedom (subsection~\ref{sec:GFMaxwellTheory}).

%%%%%%%%%%%%%%%%%%%%%%%%%%%%%%%%%%%%%%%%%%%%%%%%%%%%%%%%%%%
\subsection[General gauge in the extended Hamiltonian formalism]{General gauge in the extended Hamiltonian formalism:
\\Kinematical energy-momentum of gauge fields}\label{sec:KinEMgaugeF}

From equation~\eqref{exteom} we see that the relation
$\{ \pi_\m ,  H_E \} = \dot{\pi}_\m$ (i.e. $H_E$ generates time translations
of $\pi_\m$) does not involve $\l^1, \l^2$
and therefore holds for any value of the Lagrange multipliers.
By contrast, \emph{the relation $\{ A^\m ,  H_E \} = \dot{A}^\m$
entails that $\l^1$ and $\l^2$ are determined in terms of the basic fields}
(recall from equation~\eqref{eq:piA}
that $\vec{\pi} = \vec{E} = - \grad\, A^0 - \dot{\vec{A}}\,$):
it amounts to imposing the
  \begin{align}
   \mbox{generalized gauge conditions}
   \qquad
  \Boxed{
   \l^1 = \dot{A} ^0
  \, , \qquad
    \l^2 =0
    }
    \, .
    \label{eq:gfcond0}
    \end{align}
(Comparison with~\eqref{eq:gfcondition} shows
that the gauge condition $\l^2 =0$ is also the one which allows us to recover the Lagrangian gauge transformations from the Hamiltonian ones.)

In summary, if we impose the gauge conditions~\eqref{eq:gfcond0}, then the extended Hamiltonian~\eqref{exthamem} only depends
on the phase space variables $A^\m, \pi_\m$ (and the derivatives of $A^\m$) and reads
\begin{align}
\label{eq:kinhamcan}
 H_{\txt{kin}} \equiv
\left. H_E \right|_{\l^j \, \txt{fixed}}  =
H + \int d^3 x \, \pi^0 \dot{A} ^0
\, .
\end{align}
Here, $H$ can be decomposed into a gauge invariant part
$ H_{\txt{inv}} \equiv  \frac{1}{2}  \int d^3 x \, (\vec E ^{\, 2} + \vec B^{\, 2})$
representing the energy of the electromagnetic field,
and a remainder term, i.e.
\begin{align}
 H_{\txt{kin}} = H_{\txt{inv}} +
\int d^3 x \, \left[ \pi^0 \dot{A} ^0 + \vec{\pi} \cdot \grad\, A^0 \right]
= H_{\txt{inv}} + G_{\epsilon = A^0}
\label{kinham}
\, ,
\end{align}
where  $ G_{\epsilon = A^0}$ is the gauge generator~\eqref{eq:gaugegener}
with $\epsilon = A^0$.
\emph{By construction (see~\eqnref{eq:kinhamcan}),
the functional~\eqref{kinham} generates time translations of fields in the Hamiltonian framework:}
\begin{align}
\{ F,  H_{\txt{kin}} \} = \dot{F}
\qquad \txt{for} \ \; F\in \{ A^0, \dots, A^3, \pi_0, \dots, \pi_3 \}
\label{eq:genertime}
\, .
\end{align}

Since $H  =  \int d^3 x \, T^{00}_{\txt{can}} = P^0_{\txt{can}}$,
this line of arguments can be generalized as follows
\emph{to construct}  $P^\n _{\txt{kin}}$ \emph{with} $\{ F,  P_{\txt{kin}}^\n \} = \pa^\n {F}$.
Following reference~\cite{Hanson:1976cn}, \emph{we define extended quantities
involving Lagrange multiplier fields  $\Lambda_1^{0\n}, \Lambda_2^{0\n}$ for the FCC's:}
\begin{equation}
\label{extemt}
P_E^\n \equiv \int d^3 x \,  T_E^{0 \nu}
\equiv \int d^3 x \,
\left( T_{\txt{can}}^{0 \nu} + \Lambda_1^{0\nu}\, \pi^0 + \Lambda_2^{0\nu} \, \Div \, \vec \pi \right)
\, .
\end{equation}
We remark that notational consistency with~\eqref{exthamem} requires
$\Lambda_1^{00} = \lambda ^1$ and $\Lambda_2^{00} = \lambda ^2$, and we note that one can introduce more generally~\cite{Hanson:1976cn}
the \emph{extended EMT} within the Hamiltonian formulation by the expression
\begin{equation}
\label{eq:ExEMT}
 T_E^{\m \nu}
\equiv
 T_{\txt{can}}^{\m \nu} + \Lambda_1^{\m \nu}\, \pi^0 + \Lambda_2^{\m \nu} \, \Div \, \vec \pi
\, .
\end{equation}

With the
%extended gauge conditions
\begin{equation}
\label{lagmult}
   \mbox{extended gauge conditions}
    \qquad
\Lambda_1^{0\nu} = \partial^\nu A^0
\, , \qquad
\Lambda_2^{0\nu} = 0
\, ,
\end{equation}
which encompass the condition~\eqref{eq:gfcond0} for $\n =0$,
we then obtain the so-called \textbf{kinematical energy-momentum vector} of the gauge field,
  \begin{align}
  \label{eq:KinEMV}
 P^\n _{\txt{kin}} \equiv
\left. P^\n _E \right|_{\Lambda^{0\n} _j \, \txt{fixed}}  = P ^\n + \int d^3 x \, \pi^0 \partial^\nu A^0
\, ,
\end{align}
or
\begin{align}
  \Boxed{
 P^\n _{\txt{kin}} \equiv
\left. P^\n _E \right|_{\Lambda^{0 \n} _j \, \txt{fixed}}  = P_{\txt{inv}} ^\n + \int d^3 x \, \left[ \pi^0 \partial^\nu A^0
- A^\n \pa_i \pi^i \right]
}
\label{kinenmom}
\, .
\end{align}
Here, the last term corresponds to the last term in equation~\eqref{kinham} and the gauge invariant
contributions are the familiar ones as given in equation~\eqref{eq:gaugeinvP}.

The main result of this
% paragraph
subsection
is the following one.
By construction, \textbf{the functional \eqref{kinenmom}}
only depends on the phase space variables $A^\m, \pi_\m$ (and the derivatives of $A^\m)$
and it  \textbf{generates space-time translations
of fields without any gauge condition imposed on $(A^\m)$:}
 \begin{align}
\Boxed{
\delta_a \vp (x) \equiv
\{ \vp (x) , a_\m P^\m _{\txt{kin}} \} = a_\m \pa^\m \vp (x)
}
\qquad \txt{for} \ \; \vp \in \{ A^0, \dots, A^3, \pi_0, \dots, \pi_3 \}
 \label{kinenergymoment}
\, .
\end{align}
For $\vec a = \vec 0$, we recover expressions~\eqref{eq:kinhamcan}--\eqref{eq:genertime}.
It can be explicitly checked that we have the Poisson commutator relations
\[
\Boxed{
\{P^\m _{\txt{kin}} , P^\n _{\txt{kin}} \} =0
}
\, .
\]

On the constraint surface we have $\pi^0 =0$, hence $\Div \, \vec E =0$ by virtue of the extended
Hamiltonian equations~\eqref{exteom}.
Then $ T_{\txt{can}} ^{0\n}$ and $ T_{\txt{inv}} ^{0\n}$ differ by a divergence $\pa_i (E_{i}A^\n)$
so that the charges $P ^\n$ and  $P_{\txt{inv}} ^\n$ coincide with each other.
However, as emphasized above, the constraints must not be substituted directly in the Poisson brackets
and therefore the charges $P ^\n$, $P_{\txt{inv}}^\n$ do not generate space-time translations
of the gauge field $(A^\m)$ by virtue of the Poisson bracket:
in this respect we have to consider the charge $ P_{\txt{kin}} ^\n$ which involves an additional $\pi^0$-dependent
term, see equations~\eqref{kinenmom}, \eqref{kinenergymoment}.

We note that the decomposition of the canonical Hamiltonian $H$ into a gauge invariant part $H_{\txt{inv}}$
and a remainder term, as considered in equations~\eqref{eq:kinhamcan}, \eqref{kinham},
is also encountered in the context of the
\emph{Hamiltonian BRST quantization} where it amounts to treating the field $A^0$ as a Lagrange
multiplier in the Hamiltonian formulation~\cite{Rothe:2010}.

\paragraph{Interpretation and underpinnings of results:}
The result~\eqref{kinenergymoment} states that the infinitesimal, global, Lagrangian symmetry transformation
$\delta^{\textrm{L}}_a A^\m \equiv a_\n \pa^\n A^\m$ of the gauge field
coincides with the infinitesimal Hamiltonian symmetry transformation
 $\delta^{\textrm{H}}_a A^\m (x) \equiv \{ A^\m (x), a_\n P_{\txt{kin}} ^\n \}$.
 By virtue of~\eqref{kinenmom} and~\eqref{hamgaugesymm} we have
 \begin{align}
\label{eq:HamDecompos}
 a_\n P_{\txt{kin}} ^\n =  a_\n P_{\txt{inv}} ^\n
+  G _{(\varepsilon^1 = a\cdot \pa A^0 , \, \varepsilon^2 =-a \cdot A )}  \, ,
 \end{align}
where $  P_{\txt{inv}} ^\n $ is the gauge invariant Noether charge associated to
translation invariance of the Lagrangian action functional
and $ G _{(\varepsilon^1  , \, \varepsilon^2 )}$ represents the generator
of Hamiltonian (canonical) gauge transformations with parameters $\varepsilon^j$
depending in a specific way on the translation parameters $a^\m$ and on the gauge
field\footnote{
We thank one of the anonymous referees for his insightful comments clarifying
the nature and derivation of these results.}.
This result as well as the underlying construction~\eqref{extemt}-\eqref{kinenmom}
are quite analogous to the derivation of gauge invariant currents in the Lagrangian formulation
as presented in appendix A, e.g. compare the decomposition~\eqref{eq:HamDecompos}
with the expression~\eqref{eq:CoLieDer} of infinitesimal Lagrangian translations of gauge fields, i.e.
(for Abelian gauge fields)
\begin{align}
\delta^{\textrm{L}}_a A_\m
= \delta^{\textrm{inv}}_a A_\m + \delta^{\textrm{gauge}}_{\varepsilon =a\cdot A}  A_\m
\equiv  a^\n F_{\n \m} + \pa_\m (a \cdot A )
\, .
\end{align}

We remark that relation $\delta^{\textrm{H}}_a A_\m (x) \equiv \{ A_\m (x), a_\rho P_{\txt{kin}} ^\rho \}$
implies that
\begin{align}
\delta^{\textrm{H}}_a F_{\m \n} (x) = \{ F_{\m \n}  (x), a_\rho P_{\txt{kin}} ^\rho \}
= \{ F_{\m \n}  (x), a_\rho P_{\txt{inv}} ^\rho \}
= a^\rho \pa_\rho F_{\m \n}  (x)
\, .
\end{align}
Here, the second equality follows from the gauge invariance of the field strength tensor $F_{\m \n}$
and the last equality can be verified by using~\eqref{eq:PBAE} as well as the equations of motion, 
i.e. the free Maxwell equations. 
Thus the gauge invariant (or the canonical) energy-momentum vector generates space-time translations 
of the field strengths $\vec E$ and $\vec B$ (by contrast to the case of the gauge-variant gauge field $(A_\m)$ 
for which one has to combine this generator with the one of a specific gauge transformation).

\paragraph{Summary:}
As we noted in~\secref{sec:HamFormFT} for non singular field theories (i.e. in particular
in the absence of gauge symmetries), the Lagrangian symmetries are recovered in the Hamiltonian formulation
by the canonical Noether charges.
For gauge field theories, the Lagrangian translations of the  gauge fields $(A^\m)$ can only be recovered
by including a specific gauge transformation in the Hamiltonian symmetry variation describing translations.

More precisely, the
gauge fixing conditions~\eqref{eq:gfcond0} for the Hamiltonian gauge symmetries (i.e. for the symmetries
in extended phase space $\{ (A_\m , \pi^\m , \lambda^j )\}$  generated by the FCC's $\phi_1, \phi_2$)
allow us to reduce the Hamiltonian gauge symmetries to the usual Lagrangian gauge symmetries (i.e. to
$\delta A_\m = \pa_\m \epsilon$) and,
together with their extension~\eqref{lagmult}, they
 allow us to generate space-time translations of phase space variables,
in particular of the gauge field $(A^\m)$ in a general gauge (i.e. for  gauge potentials  $(A^\m)$
which are not constrained by any subsidiary condition).
\emph{Lorentz transformations} can be handled in a similar way, see subsection~\ref{sec:LorentzTrafo}.
We note that one may also formulate a so-called \emph{extended Lagrangian formalism}
starting from the notion of an extended configuration space and discuss symmetries in this
(less familiar) setting~\cite{DeriglazovBook2017}.
%We note that one may also formulate an ``extended'' Lagrangian in correspondence to the extended Hamiltonian, for a review of this less known %subject see \cite[Chapter 8]{DeriglazovBook2017}.

If one is interested in the \emph{quantization} of the theory, then one has to gauge fix the local symmetry $\delta A_\m = \pa_\m \epsilon$
which is at the origin of degeneracies, yielding in particular
 a singular gauge field propagator.
To implement this gauge fixing within the Hamiltonian formulation, one chooses other gauge fixing conditions than
relations~\eqref{eq:gfcond0} for $\lambda^1$ and $\lambda^2$:
We will treat this point in the next
% paragraph
subsection
where we will also come back once more
to space-time translations of the gauge field.

%%%%%%%%%%%%%%%%%%%%%%%%%%%%%%%%%%%%%%%%%%%%%%%%%%%%%%%%%%%
\subsection{Complete gauge fixing for \texorpdfstring{$(A^\m)$}{A} and Dirac brackets}\label{sec:GFMaxwellTheory}

\subsubsection{Generalities}

In order to \emph{fix the Hamiltonian gauge symmetries~\eqref{hamgaugesymm}
 generated by the FCC's} $0 \approx \phi_1 \equiv \pi ^0$ and
 $0 \approx \phi_2 \equiv \pa_i \pi ^i = \Div \, \vec{\pi}$,
 we completely break the symmetry generated by
 $G_{(\varepsilon ^j)}$
  by imposing appropriate gauge fixing conditions (such that we are only left with the two physical degrees of freedom
 of a massless vector field in four dimensions):
for each FCC $\phi_j (A, \pi )   \approx 0$, one introduces a so-called  \textbf{canonical gauge fixing condition
for the Hamiltonian gauge symmetry $G_{(\varepsilon^j)}$},
\begin{equation}
\label{gfcondi}
\Boxed{
f_j (A, \pi ) \approx 0
}
\qquad \txt{for} \ \; j\in \{ 1,2 \}
\, .
\end{equation}
The admissibility criteria for these two (independent) conditions are the usual ones, i.e.
the gauge slice $f_1= 0= f_2$ must be reachable by means of a gauge transformation and it should fix the gauge
uniquely, i.e. the gauge slice should be transversal to the gauge orbits.
Concerning the latter point, we note that the
 invertibility of the $2 \times 2$ matrix
 ${\cal A} \equiv        \left[ \{ f_j, \phi_{j^{\pr}} \} \right]$ is related to the fact that
 the only gauge transformation which leaves the gauge fixing condition $f_j = 0$
invariant is the identity transformation, i.e.
$0 = \delta_{( \varepsilon^{j^{\pr}} )} f_j = \int d^3x \, \varepsilon^{j^{\pr}} \{ f_j, \phi_{j^{\pr}} \} $ implies
$\varepsilon^{j^{\pr}} =0$ for all ${j^{\pr}}$.
Moreover, \emph{the conditions of invertibility of ${\cal A}$ and of stability of the gauge fixing condition~\eqref{gfcondi}
under time evolution imply that the Lagrange multipliers $\lambda^1, \lambda^2$ appearing in the extended
Hamiltonian are determined in a consistent manner:}
\begin{equation}
\label{eq:EqLambda}
0 \approx \dot{f} _j \approx \{ f_j, H_E \} \approx \{ f_j, H \}
+ \{ f_j, \phi_{j^{\pr}} \}\lambda ^{j^{\pr}} \, ,
\qquad \mbox{hence} \ \
\lambda^j = -  \left( {\cal A} ^{-1} \right) ^{j j^{\pr} }
 \{ f_{j^{\pr}} , H \}
 \, .
\end{equation}

Examples of admissible gauge fixing conditions for the free
electromagnetic field
are given by the \textbf{radiation gauge}
 \begin{equation}
\label{eq:radgf}
\Boxed{
\begin{array}{lcl}
\txt{FCC 1\,:} \quad
0 \approx \phi_1 \equiv \pi^0 \, , & \qquad &
\txt{Gauge fixing 1\,:} \quad
0 \approx f_1 (A, \pi ) \equiv A^0
\,,\\
\txt{FCC 2\,:} \quad
0 \approx \phi_2 \equiv \pa_i \pi^i \, , & \qquad &
\txt{Gauge fixing 2\,:} \quad
0 \approx f_2 (A, \pi ) \equiv \Div \vec{A}
\, .
 \end{array}
}
\end{equation}
or by the \textbf{special axial gauge}
 \begin{equation}
\label{eq:spaxgf}
\Boxed{
\begin{array}{lcl}
\txt{FCC 1\,:} \quad
0 \approx \phi_1 \equiv \pi^0 \, , & \qquad &
\txt{Gauge fixing 1\,:} \quad
0 \approx f_1 (A, \pi ) \equiv A^3
\,,\\
\txt{FCC 2\,:} \quad
0 \approx \phi_2 \equiv \pa_i \pi^i \, , & \qquad &
\txt{Gauge fixing 2\,:} \quad
0 \approx f_2 (A, \pi ) \equiv  \pi^3 + \pa_{3} A^0
\, .
 \end{array}
}
\end{equation}

Before considering these particular cases, we recall some generalities on the gauge fixed Hamiltonian
theory.
First, let us denote the constraints   $\phi_j$ and the corresponding gauge functions  $f_j$
collectively by  $\varphi _a$,
\begin{equation}
\label{collectcons}
\left(  \varphi_a \right) _{a= 1,\dots , 4 }
\equiv  \left(  \phi_1, \phi_2, f_1, f_2 \right)
\, ,
\end{equation}
and define the \emph{reduced phase space} $\Gamma_r$ as the  submanifold of phase space $\Gamma \equiv \left\{ \left( A , \pi \right) \right\}$ defined by the relations
$\varphi_a \approx 0$:
\begin{equation}
\Gamma_r \equiv \left\{ \left( A , \pi \right) \in \Gamma \, | \, \phi_j ( A, \pi ) =0 =  f_j ( A, \pi )
\ \; \txt{for} \ \; j= 1, 2
\right\}
\, .
\end{equation}
This space may be viewed as the \emph{physical subspace} of phase space for the constrained dynamical system under consideration.

Furthermore, we introduce the $4 \times 4$-matrix $X$ with elements $X_{ab} \equiv
\{ \varphi_a, \varphi _b \}$.
From $\{ \phi_j, \phi_{j^{\pr}} \}\! \approx 0$ (FCC's) and the invertibility of the matrix
$\left[ \{ f_j, \phi_{j^{\pr}} \} \right]$,
it follows that,
whatever the value of  the bracket $\{ f_j, f _{j^{\pr}} \} $, one has
\[
\det X =
\det \left[
\begin{array}{ccc}
\{ \phi_j, \phi _{j^{\pr}} \} & | &
\{ \phi_j, f _{j^{\pr}} \}  \\
\hline
\{ f_j, \phi _{j^{\pr}} \} & | &
\{ f_j, f _{j^{\pr}} \}
\end{array}
 \right] \approx \left( \det \left[ \{ f_j, \phi_{j^{\pr}} \} \right] \right)^2
\not\approx 0
 \, .
\]
Thus, the matrix $X$ is invertible on $\Gamma_r$:
\begin{equation}
\label{invertX}
X \approx  \left[
\begin{array}{ccc}
0 & | & -{\cal A}^t   \\
\hline
{\cal A}  & | & {\cal B}
\end{array}
 \right]
 \quad
 \Longrightarrow
\quad  X^{-1} \approx  \left[
\begin{array}{ccc}
{\cal A} ^{-1} {\cal B} ({\cal A} ^{-1})^t & | & {\cal A} ^{-1}  \\
\hline
-({\cal A} ^{-1})^t & | & 0
\end{array}
 \right]
 \, .
\end{equation}
We remark that \emph{the FCC's
$\phi _j$ supplemented with gauge fixing conditions $f_j \approx 0$
such that $\det \left[ \{ f_j, \phi_{j^{\pr}} \} \right] \not\approx 0$ can be viewed
as a set of second class constraints.}
(The fact that we do not have any FCC's anymore reflects the fact that the gauge has been completely fixed.)
The quantization
of such a \emph{purely second class system} is based on the
introduction of the so-called \textbf{Dirac bracket}:
For any two functions $F,G$ on phase space, one defines this bracket by
\begin{equation}
\label{diracbracket}
\Boxed{
\{ F,G \} _{D}
\equiv  \{ F,G \} - \{ F, \varphi_a \}
\left( X^{-1} \right) ^{ab}
\{ \varphi_b , G \}
}
 \, .
\end{equation}
The Dirac bracket enjoys the same algebraic properties as the Poisson bracket
(i.e. bilinearity, antisymmetry, the Jacobi identity, and the derivation property).
Moreover, we have
\begin{equation}
\label{dirphiF}
\{ \varphi_a  , F \} _D = 0 \qquad \mbox{for any function $F$}
\, ,
\end{equation}
since
\[
\{ \varphi_a  , F \} _D = \{ \varphi_a  , F \} - \{ \varphi_a  , \varphi_b \}
\left( X^{-1} \right) ^{bc}
\{ \varphi_c , F \} =0
\]
by virtue of $\{ \varphi_a  , \varphi_b \} = X_{ab}$.
The result (\ref{dirphiF}) means that the second class system $\left( \varphi_a \right)$
can be set to zero before or after the evaluation of the Dirac bracket. Thus, \emph{after the theory
has been formulated in terms of Dirac brackets,  the constraints and gauge fixing conditions can be used as} \textrm{strong equalities,}
\emph{i.e. as identities expressing some dynamical variables in terms of others.}
In particular, these identities can be imposed as \emph{operatorial identities}
in quantum theory where the Dirac brackets of the classical theory become commutators multiplied by $1/\ri \hbar$.

As a matter of fact, the Dirac and Poisson brackets coincide on the physical subspace $\Gamma_r$
where $\varphi_a = 0$ for all $a$.
\emph{The aim of the Dirac bracket is to eliminate the unphysical (gauge) degrees of freedom in a consistent way so as to
formulate the classical theory solely in terms of the physical degrees of freedom
using brackets which differ from the standard Poisson brackets.}
Concerning the practical determination of the inverse $X^{-1}$ and thus of the Dirac bracket~\eqref{diracbracket},
we note that one can proceed in an iterative manner by starting with a subset of the set of all constraints~\cite{Hanson:1976cn}.

We now come back again to the free Maxwell field. It
follows from the  strong equalities $\pi^0 =0$ and $\Div \vec E =0$
that the kinematical energy-momentum vector $ P^\n _{\txt{kin}}$, as defined by \eqref{eq:KinEMV}
or equivalently by~\eqref{kinenmom}, coincides with the canonical expression $ P^\n$ (or with the gauge invariant expression $ P^\n _{\txt{inv}}$) on the physical subspace.
Therefore, $ P^\n$ \textbf{generates space-time translations of the phase-space variables by means of the Dirac bracket:}
 \begin{align}
\Boxed{
\delta_a \vp (x) \equiv
\{ \vp (x) , a_\m P^\m  \}_D = a_\m \pa^\m \vp (x)
}
\qquad \txt{for} \ \; \vp \in \{ A^0, \dots, A^3, \pi_0, \dots, \pi_3 \}
 \label{eq:genSTTdir}
\, .
\end{align}

%%%%%%%%%%%%%%%%%%%%%%%%%%%%%%%%%%%%%%%%%%%%%%%%%%%%%%%%%%%
\subsubsection{Radiation gauge}
%%%%%%%%%%%%%%%%%%%%%%%%%%%%%%%%%%%%%%%%%%%%%%%%%%
We now consider our first example~\eqref{eq:radgf}  of gauge fixing conditions.
We note that the essential condition is the \emph{Coulomb gauge} choice
$\Div \vec A  \approx 0$: the Lagrangian field equation
\begin{align}
0 = \pa_\m F^{\m 0} = \pa_i F^{i 0} = \pa_i (\pa^i A^0 - \pa^0 A^i ) \approx - \Delta A^0
 \label{eq:StabCbGauge}
\, ,
\end{align}
then implies the condition $A^0  \approx 0$ for an \emph{appropriate choice of boundary condition of fields} at spatial infinity.
We mention~\cite{Das:2008} that one sometimes also considers  the \textbf{temporal gauge} choice $A^0 \approx 0$
as the basic gauge
condition in~\eqref{eq:radgf}:
the field equation $0 = \pa_\m F^{\m 0}$ then yields
$\pa_0 (\pa_i A^i)  \approx 0$. Of course $\Div \vec A  \approx 0$ represents a solution of this equation,
but in the present context there is not really a convincing argument for concluding that this represents
 the only solution since the boundary condition concerns the behavior of fields at spatial infinity.
We remark that the Coulomb gauge condition $\pa_i A^i =0$ is manifestly invariant under rotations (and under translations)
and for this reason
it was strongly advocated by J.~Schwinger (and more recently by S.~Weinberg~\cite{Weinberg:1995mt}) for the quantization of electrodynamics
and the treatment of the spin of  the photon (construction of states with good quantum numbers for momentum and angular
momentum)\footnote{The tensorial nature of the observable field strength $F_{\m \n} \equiv \pa_\m A_\n - \pa_\n A_\m$
(i.e. the transformation law $F^{\prime \m \n} (x') = \Lambda ^\m _{\ \rho}  \Lambda ^\n _{\ \sigma} F^{\rho \sigma}(x)$
for $x' = \Lambda x + a$) is not affected by assuming that Lorentz transformations of the gauge potential $A^\m$ mix with gauge transformations
(i.e. assuming that
$A^{\prime \m } (x') = \Lambda ^\m _{\ \n}  [ A^{\n }(x) + \pa ^\n \omega _{\Lambda} (x) ]$
where $ \omega _{\Lambda}$ is a real-valued function associated to the transformation $\Lambda$)
\cite{Bjorken:1965, Lautrup:1967zz, Leader:2013jra}. The Coulomb or special axial gauge conditions are not covariant if $A^\m$
is a four vector, but hold in every inertial system if $A^\m$ transforms with $\Lambda$ and an appropriately chosen function $\omega _{\Lambda}$.}.

For the radiation gauge choice~\eqref{eq:radgf},
 the matrix  ${\cal A}$ with elements
${\cal A}_{jj^{\pr}}  \equiv  \{ f_j, \phi_{j^{\pr}} \} $
appearing in (\ref{invertX}) is invertible\footnote{The inverse of
${\cal A} (\vec x, \vec y \, )$ is defined by
\[
\sum_{j^{\pr} } \int_{\br ^3} d^3 z \, {\cal A} _{j j^{\pr} } (\vec x, \vec z \, ) \, ( {\cal A} ^{-1} )_{j^{\pr} j^{\pr \pr} } (\vec z, \vec y \, )
= \delta_{j j^{\pr \pr} } \, \delta (\vec x - \vec y \, )
\, ,
\]
and it is supposed that all fields vanish at spatial infinity.}:
\begin{equation}
\label{invertA}
{\cal A} (\vec x, \vec y \, ) \approx  \left[
\begin{array}{cc}
\delta (\vec x - \vec y \, )  &  0 \\
0 &  \Delta \delta (\vec x - \vec y \, )
\end{array}
 \right]
 \quad
 \Longrightarrow
\quad  {\cal A} ^{-1}(\vec x, \vec y \, ) \approx  \left[
\begin{array}{cc}
\delta (\vec x - \vec y \, )  &  0 \\
0 & \frac{-1}{ 4 \pi} \,  \frac{1}{ |\vec x - \vec y \, |}
\end{array}
 \right]
 \, ,
\end{equation}
where we used
\begin{align}
\Delta G(\vec x - \vec y \, ) = \delta (\vec x - \vec y \, )
\qquad \txt{for} \ \;  G(\vec x - \vec y \, ) = \frac{-1}{ 4 \pi} \, \frac{1}{ |\vec x - \vec y \, |}
\, .
\label{eq:GreenFunct}
\end{align}
Thus, the Dirac brackets (\ref{diracbracket}) between the variables $A_i$ and $\pi_j$ take the form
\begin{align}
\{ A_i (t, \vec x \, ) , \pi_j(t, \vec y \, ) \}_D
  & \equiv
  \{ A_i (t, \vec x \, ) , \pi_j(t, \vec y \, ) \}
 \nonumber
\\
  &\quad
 - \int_{\br ^3} d^3 z \int_{\br ^3} d^3 w \,
 \{ A_i (t, \vec x \, ) , \varphi_a (t, \vec z \, ) \} (X^{-1})^{ab} ( \vec z, \vec w \, )
 \{  \varphi_b (t, \vec w \, ) ,  \pi_j(t, \vec y \, )  \}
 \, ,
 \nonumber
\end{align}
i.e.
\begin{align}
\Boxed{
\{ A_i (t, \vec x \, ) , \pi_j(t, \vec y \, ) \}_D
=  - \delta_{ij} \delta (\vec x - \vec y \, ) + \frac{1}{ 4 \pi} \, \pa_{x^i} \pa_{y^j}  \frac{1}{ |\vec x - \vec y \, |}
  }
  \, ,
\label{eq:APiMax}
\end{align}
or
\begin{equation}
\Boxed{
\{ A_i (t, \vec x \, ) , \pi_j(t, \vec y \, ) \}_D
=
  - \delta_{ij} ^\perp (\vec x - \vec y \, )
  }
\qquad
\mbox{with} \ \;
 \delta_{ij} ^\perp (\vec x  \, )
 \equiv
 \int_{\br ^3} \frac{d^3 k}{ (2\pi)^3} \,
\re ^{\ri\vec k \cdot \vec x  \,  }
\left(  \delta_{ij}  - \frac{k_i k_j }{ \vec k ^2 } \right)
\, ,
\label{diracbraA}
\end{equation}
where we used the fact that the inverse Fourier transform of $\vec k ^{-2}$ is $- (4 \pi |\vec x  \, |)^{-1} $.
All other fundamental Dirac brackets vanish, in particular
\begin{equation}
\label{diracbraA0}
\{ A_0 (t, \vec x \, ) , \pi^0(t, \vec y \, ) \}_D =  0
\, .
\end{equation}
Expression $\delta_{ij} ^\perp$
is known as the \emph{divergenceless} or {\em
transverse delta function} since it satisfies
$
\sum_{i=1}^3 \pa_i  \delta_{ij}^\perp (\vec x \, ) =0$ for  $j\in \{ 1,2,3 \}$.

Once the Dirac brackets are considered, the constraints and gauge fixing conditions can be used as strong equalities.
This explains why the brackets~\eqref{diracbraA}, \eqref{diracbraA0} do not have the form of canonical
(Poisson-)commutation relations.
From the physical point of view,
we are left with the vector field $\vec A$ satisfying
the wave equation $\Box \vec A = \vec 0$ and
the transversality condition $\Div\vec{A} =0$,
i.e. with the \emph{two physical degrees of freedom corresponding to
the two transverse polarizations of the photon field.}
The transformation law~\eqref{eq:genSTTdir} can be explicitly checked
with $P^\m$ given by~\eqref{eq:canonP} with
%$A^0 =0$ and
$\Div\vec E =0$,
i.e. $P^0 = \frac{1}{2} \int d^3 x \, (\vec E^ 2 + \vec B^2)$ and $\vec P  = \int d^3 x \,  E_i \vec{\nabla} A^i$
for the canonical as well as for the gauge invariant energy-momentum vectors.
In particular, we have
$\{ A_i(x) , P^j \}_D = \pa^j A_i (x)$.

Obviously, expression (\ref{diracbraA}) coincides  with the commutator
 which is obtained in quantum field
 theory by postulating canonical commutation relations
  for the creation and annihilation operators of the photon field. In fact,
\emph{the previous considerations  provide a general framework for the heuristic approach to the canonical quantization of
the free electromagnetic field (in the radiation gauge) as  considered in classic textbooks,} e.g. see reference~\cite{Bjorken:1965}.
In this context we note that the coupling of gauge fields to a source $(j^{\mu})$ consists of the addition of
a current/field coupling $- j^{\mu}A_{\mu}$ to the Lagrangian,
i.e. the addition of $+ j^{\mu}A_{\mu}$ to the Hamiltonian density. By contrast to the Lagrangian formulation
(where the equation of motion for $A^0$ in the Coulomb gauge reads $\Delta A^0 = - j^0$),
the field $A^0$ can now be set to zero consistently~\cite{Rothe:2010} since $A^0$ only amounts to a redefinition of the
Lagrange multiplier field $\lambda ^2$ in the extended Hamiltonian (\ref{exthamem}).
For a general discussion of the canonical quantization in the Coulomb gauge for matter fields coupled to the Maxwell field,
we refer to the textbook of Weinberg~\cite{Weinberg:1995mt}.

%%%%%%%%%%%%%%%%%%%%%%%%%%%%%%%%%%%%%%%%%%%%%%%%%%%%%%%%%%%
\subsubsection{Special axial gauge}
%%%%%%%%%%%%%%%%%%%%%%%%%%%%%%%%%%%%%%%%%%%%%%%%%

Next we come to our second example~\eqref{eq:spaxgf} of gauge fixing conditions.
The essential condition is the \emph{special axial gauge condition}
$A^3  \approx 0$ since the expression for the canonical momentum,
\begin{align}
\pi^3 = F^{30} = \pa^3 A^0 - \pa^0 A^3 \approx \pa^3 A^0
 \label{eq:StabAxGauge}
\, ,
\end{align}
then implies the condition $\pi^3 + \pa_3 A^0 \approx 0$.

The matrix $X$ of Poisson brackets (at fixed time $t$)
of the four constraints~\eqref{eq:spaxgf}
can easily be determined and its inverse~\eqref{invertX}
presently reads
\[
X^{-1} = \left[
\begin{array}{cccc}
0 & -g & 0 & f \\
g &  0 & f & 0  \\
0 & f & 0 & 0 \\
f & 0 & 0 & 0
\end{array}
\right]
\qquad \mbox{with} \ \; \left\{
\begin{array}{l}
\pa_{x^3} g (x,y) = f(x,y)
\\
\pa_{x^3} f (x,y) = (\pa_{x^3}) ^2 g (x,y)
= \delta (\vec x - \vec y \, )
\, .
\end{array}
\right.
\]
Thus, $g$ is a Green function of the linear differential operator $(\pa_{x^3}) ^2$
and, for an \emph{appropriate choice of boundary conditions}, the latter function is given by~\cite{Hanson:1976cn}
\begin{align}
g(x,y ) &= g (\vec x - \vec y \, ) = \frac{1}{2} \, \delta (x^1 - y^1) \, \delta (x^2 - y^2) \, | x^3 - y^3|  \,
\nn
\,, \\
\mbox{hence} \ \quad
f(x,y ) &= f (\vec x - \vec y \, ) =
\frac{1}{2} \, \delta (x^1 - y^1) \, \delta (x^2 - y^2) \, \sgn{ x^3 - y^3 }  \, .
\label{eq:GreenDel2}
\end{align}
% where $\varepsilon $ is the sign function.

From~\eqref{diracbracket} it readily follows that
the Dirac brackets of the variables $A^1, A^2, \pi^1, \pi^2$ have canonical form,
i.e we have the non-vanishing brackets
\begin{align}
\{ A_i (t, \vec x \, )  , \pi^j  (t, \vec y \, )  \}_D = \delta_i^j \delta (\vec x - \vec y \, )
\qquad \mbox{for} \ \; i,j \in \{ 1, 2 \}
\, ,
\end{align}
and by construction these variables have vanishing Dirac brackets with all constraints.

Once the Dirac brackets are considered, the fields $A^3$ and $\pi^0$ vanish while
$\pi^3 = - \pa _3 A^0$ where $A^0$ is a functional of the independent fields $(A^1, A^2, \pi^1, \pi^2)$
by virtue of the constraint $\pa_i \pi^i =0$:
Indeed, substitution of $\pi^3 = - \pa_3 A^0$
into the relation $\pa_i \pi^i =0$ yields an inhomogeneous partial
differential equation for $A^0$,
\begin{align*}
0 =  \pa_1 \pi^1 + \pa_2 \pi^2 + \pa_3 \pi^3 &= \pa_1 \pi^1 + \pa_2 \pi^2 - (\pa_3)^2 A^0
\nn\\
 &= - \pa_1 \dot{A}^1 - \pa_2 \dot{A}^2 - \Delta A^0
\, ,
\end{align*}
hence $A^0$ can be expressed in terms of $\pa_1 \pi^1 + \pa_2 \pi^2$ (or of $\pa_1 \dot{A}^1 + \pa_2 \dot{A}^2$)
by using the Green function considered for $(\pa_{x^3})^2 $ (or for   $\Delta$).
The space-time translations of the field variables are generated by the canonical Noether charges and  the
Dirac bracket.

\subsection{Lorentz transformations}\label{sec:LorentzTrafo}
%%%%%%%%%%%%%%%%%%%%%%%%%%%%%%%%%%%%%%%%%%%%%%%%%%
For  Lorentz transformations, i.e. for rotations in $\br ^3$ and for boosts,
we proceed in analogy to space-time translations.

\paragraph{Case of a general gauge:}
In this case concerning the extended Hamiltonian formalism,
we add the integral of  a linear combination of the constraint functions
$\pi^0$ and $\Div \vec E$ to the canonical charges $J^{\rho \sigma}$ (cf.~\eqnref{extemt})
\begin{align}
J^{\rho \sigma} _E \equiv J^{\rho \sigma} + \int d^3x \, \big[
\xi_1 ^{\rho \sigma} \pi^0 + \xi_2 ^{\rho \sigma} \, \Div \vec E
\, \big]
\, ,
\end{align}
and we fix the multipliers $\xi_1 ^{\rho \sigma}, \, \xi_2 ^{\rho \sigma}$ by requiring that the Poisson bracket of $A^\m $ with the functional
$J^{\rho \sigma} _E$ reproduces the correct transformation law~\eqref{eq:LtransforA} of $A^\m$.
This procedure yields the result
    \begin{align}
    \Boxed{
\delta_{\varepsilon} \vp (x) =
\{  \vp (x) , \varepsilon _{\rho \sigma} J^{\rho \sigma} _{\textrm{kin}} \}
}
\, ,
\end{align}
with (cf. equations~\eqref{eq:KinEMV},\eqref{kinenmom})
\begin{align}
 \Boxed{
 J^{ij} _{\txt{kin}}
 \equiv
\left. J^{ij} _E \right| _{ \xi_k ^{ij} \; \txt{fixed}}
 =  J^{ij} + \int d^3x \, \pi^0 \, (x^i \pa^j - x^j \pa^i ) A^0
}
\, ,
\end{align}
or
\begin{align}
\label{eq:DecompJijInv}
 J^{ij} _{\txt{kin}} =  J^{ij} _{\txt{inv}} +
 \int d^3x \, \big[ \pi^0 \, (x^i \pa^j - x^j \pa^i ) A^0
- (x^i A^j - x^j A^i)\, \Div \, \vec E  \, \big]
\, ,
\end{align}
and
\begin{align}
 \Boxed{
 J^{0i} _{\txt{kin}}
 \equiv
\left. J^{0i} _E \right| _{ \xi_k ^{0i} \; \txt{fixed}}
 =  J^{0i} + \int d^3x \, \pi^0 \, \big[ (x^0 \pa^i - x^i \pa^0 ) A^0 +A^i \, \big]
}
\, ,
\end{align}
or
\begin{align}
\label{eq:DecompJ0iInv}
 J^{0i} _{\textrm{kin}} =  J^{0i} _{\textrm{inv}} +
 \int d^3x \, \Big\{ \pi^0 \, \big[ (x^0 \pa^i - x^i \pa^0 ) A^0 +A^i \, \big]
-  (x^0 A^i - x^i A^0 ) \, \Div \, \vec E  \, \Big\}
\, .
\end{align}
We note that the gauge invariant electromagnetic field strengths  $\vec{E}$ and $\vec{B}$
have a vanishing Poisson bracket with the last terms of~\eqref{eq:DecompJijInv} and ~\eqref{eq:DecompJ0iInv}, respectively. Thus,
their Lorentz transformations are simply generated by the gauge invariant (or canonical) charges 
$J^{\rho \sigma}_{\textrm{kin}}$.

\paragraph{Case of the radiation gauge:}
In this case, it follows from $\Div \, \vec E =0$ that
 the expressions for the canonical and the gauge invariant angular momentum
vectors coincide with each other:
by virtue of~\eqref{eq:CanNClorentz} and~\eqref{eq:JoiHam}, we have the expressions
\begin{align}
J^{ij} & =  \int d^3 x \, \big[  E_{k} (x^i \pa^j - x^j \pa^i) A_k + E_{i} A^j -  E_{j} A^i \big]
\,,\nn
\\
J^{0i}
& =  \int d^3 x \, \big[ x^0 E_{k} \pa^i A_k
 - x^i \, \frac{1}{2} \, (\vec E ^{\, 2} + \vec B^{\, 2} )
\, \big]
\, .
\end{align}
One can readily verify that \emph{the rotations of the gauge field are generated by the canonical charges}
$J^{ij}$
%(with $A^0=0$ and $\Div \, \vec E =0$)
\emph{and by the Dirac bracket,} i.e. for infinitesimal rotations
with parameters $\varepsilon _{ij} = - \varepsilon _{ji}$,
we have
 \begin{align}
    \Boxed{
\delta_{\varepsilon} A^\m (x) =
\{  A^\m (x) , \varepsilon _{ij} J^{ij} \}_D
}
\, .
\end{align}
However, for a boost of $A_k$ generated by $J^{0i}$, one gets an additional contribution which has the form
of a field dependent gauge transformation~\cite{Hanson:1976cn}:
\begin{align}
\{  A_k (x) , J^{0i} \}_D =  (x^0 \pa^i - x^i \pa^0 ) A_k
- \frac{\pa \ }{\pa x^k}  \int \frac{d^3y}{4 \pi \, | \vec x - \vec y \, |} \,
 \frac{\pa A_i }{\pa x^0} (x^0 , \vec y \, )
\, .
\end{align}
In fact, the latter term ensures the vanishing of the bracket
$ \{ \pa_k A_k (x) , J^{0i} \}_D $ which has to hold by virtue of the gauge fixing condition $\Div \, \vec A=0$
(which is not invariant under Lorentz boosts).
Quite generally, for an \emph{infinitesimal boost of the gauge field} $(A^\m )$ parametrized by $\varepsilon _{0i}$,
we  have to include (for the radiation gauge)
 the field dependent gauge transformation appearing in the previous equation:
\begin{align}
A ^{\prime \m} (x) = A^\m (x) +   \varepsilon _{0i} \left[ (x^0 \pa^i - x^i \pa^0 ) A^\m
+ \eta^{\m 0} A^i - \eta^{\m i } A^0
- \frac{\pa \ }{\pa x^\m}  \int \frac{d^3y}{4 \pi \, | \vec x - \vec y \, |} \,
 \frac{\pa A_i }{\pa x^0} (x^0 , \vec y \, ) \right]
 .
\end{align}
The gauge condition $A ^{\prime 0} =0$ then also holds by virtue of the field equations
(which read $\Box A^i=0$ in the radiation gauge).

%%%%%%%%%%%%%%%%%%%%%%%%%%%%%%%%%%%%%%%%%%%%%%%%%%%%%%%%%%%%%%%%%%%%%%%%%%%%%%%%%%%%%%%%%%%
\section{On the quantization of Abelian gauge field theory}\label{sec:AssessMax}

The \emph{canonical quantization} (operator quantization)
of a classical constrained Hamiltonian system
consists of replacing the Dirac brackets by $1/(\ri \hbar )$ times the commutators of
the corresponding operators. In this section, we have another look at the derivation of Dirac brackets and we put the considered approach
to quantization into a general context.
Concerning the choice of gauges, we note that a given choice
 will be more or less convenient depending on the problem under consideration.
 The form or derivation of the Poincar\'e transformations will also be commented upon for the different formulations.

%%%%%%%%%%%%%%%%%%%%%%%%%%%%%%%%%%%%%%%%%%%%%%%%%%%%%%%%%%%%%%%%%%%%%%%%%%%%%%%%%%%%%%%%%%%
\subsection{Canonical quantization with a complete gauge fixing}
Dirac's approach to constrained Hamiltonian systems starts with the primary constraints which
result directly from the Lagrangian without any reference to the equations of motion:
the stability condition for the primary constraints
 (i.e. their preservation under the time evolution defined by the primary Hamiltonian $H_{p}$)
may yield a secondary constraint whose stability may lead to a tertiary constraint and so on. Thus, one has a \emph{chain of constraints}
(which stops in practice after a few steps), e.g. we have a total of two constraints for the free Maxwell theory.
FCC's correspond to local Hamiltonian symmetries which have to be gauge fixed so as to eliminate the redundant degrees of freedom.
To realize the gauge fixing, one can proceed as for the constraints, i.e. one imposes a single gauge fixing condition and then determines the equations which follow from it
by imposing its preservation under time evolution, while iterating the procedure for the resulting equation.
Thereby one obtains a \emph{chain of gauge fixing conditions}~\cite{Burnel:2008zz}.
By proceeding along these lines for free Maxwell theory, we found in equations~\eqref{eq:StabCbGauge}
and~\eqref{eq:StabAxGauge} that the Coulomb gauge fixing condition $\Div \vec A  \approx 0$
yields $A^0 \approx 0$ (upon a proper choice of boundary condition of fields at spatial infinity)
and that the special axial gauge condition $A^3 \approx 0$ yields $\pi^3 + \pa_3 A^0  \approx 0$.
Stability of these ``secondary gauge fixing conditions'' yields an equation which fixes the undetermined
primary Lagrange multiplier $\lambda ^1$. Thus, for the Coulomb and special axial gauge fixing conditions,
one has as many independent gauge fixing conditions as constraints: the gauge is fixed completely
(which implies that the four degrees of freedom of the gauge field $(A^\m )$ are reduced to its two physical degrees of freedom).
This type of gauge fixing is referred to as \emph{class I gauge fixing} in the terminology of Burnel~\cite{Burnel:1982fb, Burnel:2008zz}.
The fact that the Lorentz invariance is not realized manifestly in this approach is unpleasant for calculations, but does not raise a problem
for the final physical results since the latter can be shown to be Lorentz invariant (eventually with a fair amount of labor,
see~\cite{Manoukian:1987hy} and references therein).
We note that apart from the radiation gauge and the special axial gauge there exist some other interesting  complete gauge fixing
conditions, in particular the so-called \emph{light-cone} or \emph{light-front gauge}~\cite{Hanson:1976cn}.

Eventually, one may also try to \emph{solve explicitly the constraints,} e.g. for the radiation gauge by decomposing the fields
into transversal and longitudinal components and then investigating the brackets between the latter:
This approach has some advantages, but it involves non-local expressions
and does not strictly follow the canonical procedure in that it ignores the conjugate momentum $\pi^0$~\cite{Hanson:1976cn}.
We will briefly expand
on this approach in equation~\eqref{eq:Helmholtz} below.

A general issue of the approach of complete gauge fixing is the unavoidable occurrence of non-localities.
For instance, the implementation of the radiation gauge results in Dirac brackets involving a non-local term (i.e.
the second, derivative term on the right hand side of equation~\eqref{eq:APiMax}). Similarly, for the special axial gauge, the variable
$A^0$ depends on the independent variables $(A^1, A^2, \pi^1, \pi^2 )$ by means of an integral, i.e. a non-local expression.
These \emph{non-localities appearing for a complete gauge fixing}
result from the derivative terms in the gauge fixing conditions ($\Div \vec A  \approx 0$ and   $\pi^3 + \pa_3 A^0  \approx 0$,
respectively) and can be traced back to the presence of derivatives in the constraint $\pa_i \pi^i \approx 0$.
These non-localities in the Hamiltonian formulation of Abelian gauge field theory do not represent an obstacle
for investigating the corresponding quantum field theory and for deriving
important physical results~\cite{Bjorken:1965,Weinberg:1995mt}.
However, the whole framework is not fully compatible with the axioms of local relativistic field theory.

At this stage, we also mention the alternative \emph{approach to constrained dynamical systems
proposed by  L.~D.~Faddeev and R.~Jackiw}~\cite{Faddeev:1988qp, Jackiw:1993in, Rothe:2010}
which is essentially equivalent to Dirac's procedure~\cite{\Garcia}.
The idea of this approach is to formulate the theory in canonical
form on reduced phase space, i.e. solely in terms of unconstrained variables which describe the physical degrees of freedom.
Yet, the determination of the reduced coordinates amounts to solving explicitly the FCC's and gauge fixing conditions
of Dirac's approach and thereby non-local expressions appear for the basic variables in
gauge field theories~\cite{Jackiw:1993in, Rothe:2010}.

%%%%%%%%%%%%%%%%%%%%%%%%%%%%%%%%%%%%%%%%%%%%%%%%%%%%%%%%%%%%%%%%%%%%%%%%%%%%%%%%%%%%%%%%%%%%
\subsection[Canonical quantization in the Lorenz gauge]{Canonical quantization in the Lorenz gauge $\pa_\m A^\m \!=0$ (Gupta-Bleuler method)}
An alternative to complete gauge fixing within the Hamiltonian formulation consists of modifying the initial Lagrangian
by adding to it a \emph{Lorentz covariant gauge fixing term,} e.g. involving $(\pa_\m A^\m)^2$: this implies that the field $\pi^0$ no longer vanishes
and that one has a time evolution equation for all components $A^\m$ of the gauge field.
The historical realization of this idea (which goes back to W.~Heisenberg in 1928) is to consider ${\cal L} \leadsto
{\cal L} + {\cal L}_{\textrm{fix}}(A) $
with ${\cal L}_{\textrm{fix}}(A) \equiv - \frac{1}{2\xi} \, (\pa_\m A^\m)^2$
where $\xi$ is  a real non-zero parameter, a convenient choice being $\xi =1$ (``Feynman gauge'').
 The gauge field is now unconstrained and involves four degrees of freedom which
 describe two transverse polarizations, a longitudinal one and a scalar one.
 Accordingly, the Hilbert space of states in quantum theory involves more states than just the physical ones.
 In the classical Lagrangian field theory,
 we have (for $\xi =1$) the equation of motion $\Box A^\m =0$ which implies $\Box (\pa_\m A^\m) =0$.
 Thus, $\pa_\m A^\m $ represents a free scalar field which can eventually be put to zero,
 thus implementing the \emph{Lorenz gauge  condition} (L.~Lorenz, 1867). However, in the quantum theory, the Lorenz gauge condition cannot be imposed as an operatorial
 identity $\pa_\m A^\m =0$ since the latter is inconsistent with the canonical commutation relations for $A^\m$ and $\pi^\n$.
 The way out (i.e. the method to reduce the number of degrees of freedom to the physical ones)
 is based on a proposal by E.~Fermi (1929) and consists of imposing a weaker condition on the theory
 by restricting the full state space  ${\cal H}$ to the subspace ${\cal H}_{\textrm{phys}}$ of vectors $|\Psi \rangle$
 for which the gauge constraint is satisfied in the mean, i.e.
 $\langle \Psi | \pa_\m A^\m |\Psi \rangle =0$.
 The successful implementation of this program was put forward in 1950
 (in the Feynman gauge)
 by S.~N.~Gupta for the free field case
 and by K.~Bleuler for the interaction of the radiation field with matter~\cite{Das:2008}: Fermi's condition is realized for states $|\Psi \rangle$
 which satisfy the
 \begin{align}
 \mbox{\textbf{Gupta-Bleuler subsidiary condition:}}
 \qquad
 \pa^\m A^{(+)} _\m (x) \, |\Psi \rangle =0
 \qquad \mbox{for all} \ \; x
 \, ,
 \label{eq:GBsc}
 \end{align}
 where
 $ A^{(+)} _\m (x) \equiv \int \frac{d^3k}{(2 \pi )^{3/2} \sqrt{2 |\vec k \, |}} \, a_{\m} ( |\vec k \, |, \vec k) \, \re^{-\ri kx}$
 represents the positive frequency part of $A^\m $.
 Thus, \emph{the gauge field is unconstrained,} but the state space is restricted to a subspace.
 (One also says that the Lorenz condition holds in the mean for certain states.)
 Since the Poisson brackets for the gauge field $(A^\m )$ and its conjugate momentum $(\pi_\n )$ have the canonical form,
 the \emph{Poincar\'e transformations} of fields are generated by these brackets in the standard manner.

%%%%%%%%%%%%%%%%%%%%%%%%%%%%%%%%%%%%%%%%%%%%%%%%%%%%%%%%%%%%%%%%%%%%%%%%%%%%%%%%%%%%%%%%%%%%
\subsection{Canonical quantization by a generalized Gupta-Bleuler procedure}
The Gupta-Bleuler approach  for the Lorenz gauge described above can also be formulated by
introducing a scalar Lagrange multiplier field $b$ and considering the following modification of the gauge invariant Maxwell field Lagrangian
(the modification in this general form being  due to T.~Kibble~\cite{Kibble:1967sv}):
\begin{align}
\label{eq:KibbleLag}
{\cal L} \leadsto {\cal L} + {\cal L}_{\textrm{fix}}(A,b ) \, ,
\qquad \mbox{with} \quad
{\cal L}_{\textrm{fix}}(A,b ) \equiv  b \, (\pa_\m A^\m) + \frac{\xi}{2} \, b^2
\, ,
\end{align}
 where $\xi$ is an arbitrary (possibly zero) real constant.
The equation of motion of $b$
 (i.e. $b = -\frac{1}{\xi} \, \pa_\m A^\m$ if $\xi \neq 0$) then states that the scalar field $b$ coincides up to a factor with the
 field $\pa_\m A^\m$. If one substitutes this
  equation into the Lagrangian ${\cal L}_{\textrm{fix}}(A,b ) $ then one recovers the Lagrangian
  ${\cal L}_{\textrm{fix}}(A )= - \frac{1}{2\xi} \, (\pa_\m A^\m)^2$
  which is usually considered in the Gupta-Bleuler approach (with $\xi =1$). We remark that one may refer to  $ b (\pa_\m A^\m) + \frac{\xi}{2} \, b^2$ as the
  first order form of the gauge fixing Lagrangian and to $- \frac{1}{2\xi} \, (\pa_\m A^\m)^2$
  as the second order form~\cite{Kibble:1967sv} very much like the first and second order forms of the gravitational Lagrangian.

The $b$-field formulation~\cite{Burnel:2008zz} which was put forward by
  N.~Nakanishi~\cite{Nakanishi:1966zz} and B.~Lautrup~\cite{Lautrup:1967zz}
 towards 1966 (with some later refinements~\cite{Nakanishi:1972pt})
represents an elegant generalization of the Gupta-Bleuler approach
to the case of generic values of the gauge parameter $\xi$.
In this formulation,
the application of $\pa_\n$ to the equation of motion of $A_\m$ (i.e. to $\pa_\m F^{\m \n} = \pa^\n b$) yields
$\Box b =0$, i.e. $b$ represents a free scalar field.
 The \emph{Gupta-Bleuler subsidiary condition} presently becomes the
  \begin{align}
 \mbox{\textbf{Nakanishi-Lautrup subsidiary condition:}}
 \qquad
 b^{(+)} (x) \, |\Psi \rangle =0
 \qquad \mbox{for all} \ \; x
 \, ,
 \label{eq:NLsc}
 \end{align}
 which   ensures that longitudinal
 and scalar photons do not contribute to physical processes~\cite{Burnel:2008zz}.
By virtue of the equation of motion of $b$, i.e. $b = -\frac{1}{\xi} \, \pa_\m A^\m$, the condition~\eqref{eq:NLsc}
is equivalent (for $\xi \neq 0$) to the
 Gupta-Bleuler condition~\eqref{eq:GBsc}.
 For the application of Dirac's Hamiltonian approach to the modified (gauge fixed) Maxwell Lagrangian~\eqref{eq:KibbleLag},
 we refer to the works~\cite{Burnel:2008zz, Kugo}.

It turns out that this approach based on the introduction of an auxiliary field can be generalized
to much more general \emph{linear gauge fixing conditions} than the Lorenz gauge, in particular to \textbf{algebraic non-covariant gauges}
and to gauges interpolating between various of these gauges, see~\cite{Burnel:2008zz}
and references therein. Here, we only spell out the \emph{gauge fixing Lagrangian} and a few particular cases which are covered by the latter:
\begin{align}
\label{eq:BurnelLagr}
{\cal L} _{\textrm{fix}}(A,b, b' )= -C_{\m \n} (\pa^\m b)  A^\n  + \frac{\xi}{2} \, b^2
 + \frac{\xi '}{2} \,  (\pa^\m b') (\pa_\m b') + \xi ' \, b \, C^\m \pa_\m b'
 \, .
\end{align}
In this expression, $b$ and $b'$ are two independent real scalar fields, $\xi$ and $\xi '$ two independent real gauge parameters,
$C_{\m\n}$ is a given constant, not necessarily symmetric, tensor of rank two (with $C_{00} \neq 0$), and $C_\m$ a given constant four-vector.
Interesting particular cases are obtained by expressing $C_{\m\n}$ and $C_{\m}$ in terms of the Minkowski metric $\eta_{\m \n}$
and some fixed four-vectors $n, n^*$.
The equation of motion of the auxiliary field $b$ yields the
%\emph{gauge fixing condition}
\begin{align}
\label{eq:BurnelGFC}
\mbox{gauge fixing condition}
\qquad
0 = C_{\m \n} (\pa^\m A^\n ) + {\xi} \, b
 + {\xi '} \,  C^\m \pa_\m b'
 \, .
\end{align}
For instance, for  $C_{\m\n} =\eta_{\m \n}$ and $\xi '=0$, we recover the Lorenz gauge condition discussed above. In this case,
the relativistic invariance is manifestly realized.
Another interesting particular case is given by the choice  $\xi =0 =\xi '$ and $C_{\m\n} = n_\m n_\n - \alpha \eta_{\m \n}$ where
$(n_\m)$ is a fixed four-vector (with $n^2 >0$) and $\alpha$ a real constant:
The gauge fixing condition~\eqref{eq:BurnelGFC} then reads $0= (n\cdot \pa ) (n\cdot A) - \alpha \pa \cdot A $,
hence  for $\alpha \to \infty$ we recover the Lorenz gauge and
for $\alpha =1$ we have a condition generalizing the Coulomb gauge choice.
Indeed, the latter is realized in the special frame where $n=(1, \vec 0 \, )$
so that $(n\cdot \pa ) (n\cdot A) - \pa \cdot A = - \Div \, \vec A$.
Remarkably, with some amount of labor~\cite{Burnel:2008zz}, the Gupta-Bleuler procedure can be applied for the general gauge fixing Lagrangian~\eqref{eq:BurnelLagr}.
More precisely,  the subsidiary condition selecting physical states $|\Psi \rangle $ that we encountered above, i.e.
$b^{(+)}(x) \,  |\Psi \rangle =0$, now has to be supplemented
together with the same condition involving the auxiliary field $b'$.

For $C_{\m\n} \neq \eta_{\m \n}$ or $C_\m \neq 0$,
\emph{Lorentz invariance} is broken in the classical theory due to the presence of these fixed tensors, hence
the total Lagrangian no longer transforms like a scalar
 under the infinitesimal Lorentz transformations~\eqref{eq:LtransforA}. It rather transforms as $\delta {\cal L}
 = \varepsilon_{\m \n} (x^\m \pa^\n - x^\n \pa^\m ) {\cal L} +K^{\m \n} - K^{\n \m}$ where $K^{\m \n}$ is an asymmetric tensor depending on
$A^\m , b$ and $b'$ (which vanishes for $C_{\m\n} = \eta_{\m \n}, \, C_\m = 0$).
Consequently, the canonical angular momentum tensor, as given in
 equation~\eqref{eq:MaxAMT}, is  not conserved:
 $\pa_\m M_{\textrm{can}} ^{\mu \rho \sigma} = K^{\rho \sigma} -  K^{\sigma \rho}$. However, in quantum theory,
 the subsidiary  conditions $b^{(+)}(x) \,  |\Psi \rangle =0 = b'^{(+)}(x) \,  |\Psi \rangle =0$ imply that the
 expectation values of the operators $b, b'$ vanish for physical states and thereby  the (normally ordered)
 operator $K^{\m \n}$ also does, i.e. $\langle \Psi |  : \! \! K^{\m \n} \! \! : | \Psi \rangle =0$. Henceforth,
 \emph{Poincar\'e invariance} holds in the physical
 sector of the underlying quantum field theory~\cite{Burnel:2008zz}.

%%%%%%%%%%%%%%%%%%%%%%%%%%%%%%%%%%%%%%%%%%%%%%%%%%%%%%%%%%%%%%%%%%%%%%%%%%%%%%%%%%%%%%%%%%%%
\subsection{BRST quantization and path integral  quantization}
 A powerful generalization of the Gupta-Bleuler approach to Abelian gauge field theory
 is given by the BRST quantization.
 The latter also allows to tackle
 non-Abelian (i.e. non-linear) gauge field theories for which the Gupta-Bleuler procedure no longer works.
 It can be applied within the Lagrangian or the Hamiltonian formulation of field theory
 and it allows us to implement a large variety of Lorentz covariant or non-covariant gauge choices,
 including  \emph{non-linear} ones:
 We will discuss
 this point further in section~\ref{sec:AssessYM} (where we also comment on the path integral
 approach and on the relationships between these different approaches).
 Here we only note that \emph{the BRST quantization
 results  in the Hamiltonian framework in a characterization
 of physical states as those which are left invariant by the so-called BRST operator:}
 For Abelian gauge field theory, the latter condition is nothing else but the Gupta-Bleuler subsidiary condition,
 see equations~\eqref{eq:TotAct}--\eqref{eq:BRSTinvStates} below.
 An example for a \emph{non-linear gauge} in electrodynamics is given by the \emph{`t Hooft-Veltman gauge}~\cite{tHooft:1972ue}
 i.e. $0= \pa_\m A^\m + \frac{1}{2} \alpha A_\m A^\m$,
 where $\alpha \neq 0$ represents a real dimensionless constant (see references~\cite{Joglekar:1974xk, Rouet:1975xg, Mckeon:1986zy}
 for a study of this gauge).

%%%%%%%%%%%%%%%%%%%%%%%%%%%%%%%%%%%%%%%%%%%%%%%%%%%%%%%%%%%%%%%%%%%%%%%%%%%%%
%%%%%%%%%%%%%%%%%%%%%%%%%%%%%%%%%%%%%%%%%%%%%%%%%%%%%%%%%%%%%%%%%%%%%%%%%%%
\section{Hamiltonian formulation of pure non-Abelian YM theory}
\label{sec:HamYM}

In this section, we outline the non-Abelian generalization
of the results presented in section~\ref{sec:HamFormMaxwell} concerning the free Maxwell
theory in four dimensions.
Thus, our starting point is the action functional~\eqref{eq:invact}.

%%%%%%%%%%%%%%%%%%%%%%%%%%%%%%%%%%%%%%%%%%%%%%%%%%%%%%%%%%%%%%%%%%%%%%%%%%%%%%%%%%%%%%%%%%%%%%%%%%%%%%%%%%%
\subsection{Canonical momenta and Hamiltonian}

With the notation $\pi^\m \equiv {\pa {\cal L}}/ {\pa \dot{A} _\m} =F^{\m 0}$
and $  F^{i0} \equiv E_i, \, F^{ij} \equiv -  \varepsilon ^{ijk} B_k$
(chromo-electric and -magnetic fields),
 the  conserved charges
$P^\n \equiv \int d^3 x \, T_{\txt{can}}^{0 \nu}$ following from the local conservation law $\pa_\m T_{\txt{can}}^{\mu \nu} =0$
have the form~\eqref{eq:canonP}, \eqref{eq:gaugeinvP}:
\begin{align}
P^0 \equiv H  & =  H_{\textrm{inv}} +
\int_{\br ^3} d^3 x \,  \Tr \, [  - A^0 (D_i \pi^i ) ] \, ,
 \qquad \mbox{with} \ \;
 H_{\textrm{inv}} \equiv \int_{\br ^3} d^3 x \,  \Tr \, \Big[  \frac12 \, ( \vec E^{\, 2} + \vec B^{\, 2} ) \Big]
 \, ,
 \nn
 \\
 \vec P  & = \vec P _{\textrm{inv}} +
\int_{\br ^3} d^3 x \,  \Tr \, [  - \vec A \, (D_j \pi^j ) ] \, ,
 \qquad \ \, \mbox{with} \ \;
 \vec P _{\textrm{inv}} \equiv \int_{\br ^3} d^3 x \,  \Tr \, ( \vec E \times \vec B \, )
 \, .
 \label{eq:TransGenYM}
\end{align}
One readily finds that
the constraints~\eqref{primconsem}, \eqref{gauss} of the free Abelian gauge theory
presently generalize to Lie algebra-valued  constraints
\begin{align}
\Boxed{
\pi^0 \approx 0 \, , \qquad D_i \pi ^i \approx 0
}
 \, ,
 \label{eq:YMcons}
\end{align}
which are again of first class.

%%%%%%%%%%%%%%%%%%%%%%%%%%%%%%%%%%%%%%%%%%%%%%%%%%%%%%%%%%%%%%%%%%%%%%%%%%%%%%%%%%%%%%%%%%%%%%%%%%%%%%%%%%%
\subsection{General gauge: kinematical energy-momentum of gauge fields}

The Abelian gauge theory expressions~\eqref{eq:ExEMT}--\eqref{kinenmom}
now generalize to
\begin{equation}
\label{eq:YMExEMT}
 T_E^{\m \nu}
\equiv
 T_{\txt{can}}^{\m \nu} +  \Tr \, [ \Lambda_1^{\m \nu}\, \pi^0 + \Lambda_2^{\m \nu} \, D_i \pi^i  ]
\, , \qquad \mbox{with} \ \; \Lambda_1^{0\nu} = \partial^\nu A^0
\, , \quad
\Lambda_2^{0\nu} = 0
\, ,
\end{equation}
hence the \textbf{kinematical energy-momentum vector of gauge fields} is given by
\begin{align}
  \Boxed{
 P^\n _{\txt{kin}} \equiv
\left. P^\n _E \right|_{\Lambda^{0 \n} _j \, \txt{fixed}}
= P_{\txt{inv}} ^\n + \int d^3 x \,  \Tr \, \left[ \pi^0 \partial^\nu A^0
- A^\n D_i \pi^i \right]
}
\label{eq:YMkinenmom}
\, ,
\end{align}
where the
contributions $ P_{\txt{inv}} ^\n$ are the ones specified in equation~\eqref{eq:TransGenYM}.

%%%%%%%%%%%%%%%%%%%%%%%%%%%%%%%%%%%%%%%%%%%%%%%%%%%%%%%%%%%%%%%%%%%%%%%%%%%%%%%%%%%%%%%%%%%%%%%%%%%%%%%%%%%%%
\subsection{Results for the radiation gauge}

Let us consider the \emph{Coulomb gauge} fixing condition
\begin{align}
  \Boxed{
\Div \vec A \approx 0
}
\, , \qquad \mbox{i.e.} \ \;
\pa_i A^i _a \approx 0 \ \ \mbox{for} \ \; a \in \{ 1, \dots , n_G \}
\, .
\label{eq:YMCb}
\end{align}
From the expression  of the canonical momentum,
\begin{align}
\pi^i = F^{i0} = \pa ^i A^0 - \pa^0 A^i + \ri q [ A^i , A^0]
\, ,
\label{eq:PiF}
\end{align}
it follows by substitution of the  Coulomb gauge condition for $\vec A$ that
we have a partial differential equation for $A^0$:
\[
\pa_i \pi^i \approx  -\Delta A^0 + \ri q [ A^i ,\pa_i A^0]
\, ,
\]
i.e.
\begin{align}
 Q^{ab} A^0_b \approx - \pa^i \pi^a _i
\, ,
 \qquad
\mbox{with} \ \;
Q^{ab} \equiv \delta^{ab} \Delta - q f^{abd} A^i_d \pa_i
\, .
\label{eq:IDE}
\end{align}
Here, the linear differential operator $Q^{ab}$ represents a deformation of the Laplacian operator
which is parametrized by the potential $\vec A$.
We note that, by virtue of the  secondary constraint  $D_i \pi ^i \approx 0$,
 the divergence of $\vec{\pi}$ may also be written as a commutator:
$ \pa_i \pi^i \approx - \ri q  [ A_i ,\pi^i]$.
(As a matter of fact, the latter commutator  represents
the density of the conserved charge which is associated to the invariance of the action functional under global gauge transformations.)

A solution $A^0$ of the  inhomogeneous differential equation~\eqref{eq:IDE} is obtained by convoluting an inverse $G_{ab}$ of $Q^{ab}$
(i.e. a Green function of the differential operator $Q^{ab}$)
with the inhomogeneous term $ - \pa^i \pi^a _i$ (or equivalently with
$\ri q [ A^i ,\pi_i ]^a $):
\begin{align}
  \Boxed{
A^0_a (x) + \int d^3 y \, G_{ab} (x,y) \, \pa_{y^i} \pi^i_b (y) \approx 0
}
\, .
\label{eq:YMCbcons}
\end{align}
More precisely, we consider
 the $\vec A$-dependent Green function $G_{ab} (x, y)$ defined by
 \begin{align}
 Q^{ab}G_{bc} (x,y) = \delta^a_c \, \delta (\vec x - \vec y \, )
 \, ,
\label{eq:YMGreen}
\end{align}
 which decays as $1/r$ at spatial infinity.
Although one does not have an explicit expression for $G_{ab}$,
relation~\eqref{eq:YMGreen} can be solved iteratively and thus $G_{ab}$
can be written~\cite{Hanson:1976cn} as a deformation of the Green function~\eqref{eq:GreenFunct}
of the Abelian theory (which decays as $1/r$), namely as $\delta_{ab} G $ plus an infinite power series in the coupling constant $q$.

In summary, we have the \emph{constraint equations}~\eqref{eq:YMcons} and the
\emph{gauge fixing conditions}~\eqref{eq:YMCb} and~\eqref{eq:YMCbcons}, i.e.
a set of relations which reduces to~\eqref{eq:radgf} for the Abelian theory.
Determination of the Dirac brackets leads to a non-linear generalization of the bracket~\eqref{eq:APiMax},
\begin{align}
\label{eq:YMradDir}
\Boxed{
\{ A_i^a (t, \vec x \, ) , \pi_j^b (t, \vec y \, ) \}_D
 =
  - \delta ^{ab} \delta_{ij} \delta (\vec x - \vec y \, ) - D_i^{ac}  \pa_{y^j}  G_{cb} (x, y)
  }
\, ,
\end{align}
where $D_i^{ac} = \delta^{ac}  \pa_{x^i} - q f^{abc} A^b_i (x)$ represents the covariant derivative.
(By construction, the bracket~\eqref{eq:YMradDir} is compatible with the constraints $\pa^i A_i^a =0$
and $D^j \pi_j^b =0$.)
One also finds  non-vanishing
Dirac brackets for
%%% (see Roma lectures' eq. (6.48))
$A^0_a$ and $A^0_b$, for  $A^0_a$ and $A^i_b$
%$A^0_a$ and $\pi^0_b$,
and for $\pi_i^a $ and $\pi_j^b $.
By contrast to the Abelian theory, the Dirac brackets are now highly non-local in the gauge field $\vec A$
(so that the quantization becomes an extremely difficult endeavor).
If one considers the Dirac brackets rather than the Poisson brackets, all constraints can be imposed as strong equalities and the kinematical
energy-momentum four-vector~\eqref{eq:YMkinenmom} then reduces to the expression $P^\n _{\textrm{inv}}$ specified in equation~\eqref{eq:TransGenYM}.
The generators  of \emph{Lorentz transformations} can be discussed along the same lines.

We note that~\eqref{eq:YMradDir} are not the commutators which are generally considered in the literature
for the quantization of YM theories in the Coulomb gauge~\cite{Christ:1980ku,Zwanziger:1998ez,Rocha:2009xq}
where one rather decomposes the Lie algebra-valued canonical momentum $\vec{\pi} \equiv (\pi ^i)$ into transverse and longitudinal parts,
i.e. $\vec{\pi} = \vec{\pi}_{\perp} - \vec{\nabla }\Omega$.
By virtue of the Coulomb gauge condition $\Div \vec A =0$, the constraint equation $0= D_i \pi^i = \vec D \cdot \vec{\pi}$
is then equivalent to the transversality condition $\Div \, \vec{\pi} _{\perp} =0$
(as in the Abelian case) supplemented with the condition
\[
\vec{\nabla } \cdot \vec D \, \Omega = - \rho \, ,
\qquad \mbox{with} \ \;
\rho \equiv - \ri q \, [ A_i, \pi_{\perp} ^i ] \, .
\]
Very much like~\eqref{eq:IDE} with $ \pa_i \pi^i \approx - \ri q  [ A_i ,\pi^i]$,
this relation represents the non-Abelian generalization of the Poisson equation of electrodynamics with $\rho$ being
interpreted as the \emph{density of color charges}  of the gauge fields.
For the components of $\vec A$ and $\vec{\pi} _{\perp}$, one is then led to
 a commutator having the same form as in the Abelian theory, i.e. expression~\eqref{eq:APiMax}.
The corresponding expression for the generators of \emph{Poincar\'e
transformations} and the relativistic invariance of the Hamiltonian formulation of YM in the Coulomb gauge
 are discussed in references~\cite{Besting:1989nq, Rocha:2009xq}.

%%%%%%%%%%%%%%%%%%%%%%%%%%%%%%%%%%%%%%%%%%%%%%%%%%%%%%%%%%%%%%%%%%%%%%%%%%%%%%%%%%%%%%%%%%%%%%%%%%%%
\subsection{Results for the special axial gauge}

For the \emph{special axial gauge} condition $A^3 \approx 0$, equation~\eqref{eq:PiF}
implies $\pi^3 = \pa^3 A^0$, hence we have the \emph{gauge fixing} conditions
\begin{align}
  \Boxed{
A^3 \approx 0 \, , \qquad \pi^3 + \pa_3 A^0 \approx 0
}
\, ,
\label{eq:YMAG}
\end{align}
i.e. we have $n_G$ copies of the gauge fixing condition~\eqref{eq:spaxgf}
 discussed for Abelian gauge theory (though the constraint
equations~\eqref{eq:YMcons} presently involve the covariant derivative of $\vec{\pi}$).
Thus, there are close parallels with the Abelian theory: the inverse $X^{-1}$ of the matrix $X$
of Poisson brackets again involves the Green function $g$ of the operator $(\pa_{x^3})^2$
as given by~\eqref{eq:GreenDel2},
and we again have the four independent variables $(A^1, A^2, \pi^1, \pi^2 )$
(which are now Lie algebra-valued) and whose Dirac brackets have the canonical form.
By virtue of the constraints and gauge fixing conditions, the variables $A^3$ and $\pi^0$ vanish
while $A^0$ and $\pi^3$
can be expressed in terms of the independent variables $(A^1, A^2, \pi^1, \pi^2 )$
by means of the Green function $g$.
The \emph{space-time translations} are once more generated by the canonical Noether charges and the Dirac brackets.

%%%%%%%%%%%%%%%%%%%%%%%%%%%%%%%%%%%%%%%%%%%%%%%%%%%%%%%%%%%%%%%%%%%%%%%%%%%%%%%%%%%%%%%%%%%%%%%%%%%%
\section{On the quantization of non-Abelian gauge field theory}\label{sec:AssessYM}

%%%%%%%%%%%%%%%%%%%%%%%%%%%%%%%%%%%%%%%%%%%%%%%%%%%%%%%%%%%%%%%%%%%%%%%%%%%%%%%%%%%%%%%%%%%
\subsection{Canonical quantization with a complete gauge fixing}
The remarks made in section~\ref{sec:AssessMax} concerning the \emph{canonical quantization}
of Abelian gauge field theory also hold for the non-Abelian case:
interesting physical results can be derived, but this  approach lacks manifest Lorentz invariance
and it involves non-local terms. The latter problem is presently worsened quite severely due to the
complicated (non-polynomial) field dependence of the non-local terms, e.g. the last term in the Dirac brackets~\eqref{eq:YMradDir}
or in relation~\eqref{eq:YMCbcons}.
For instance, a proof of renormalizability of YM-theory
 in the Coulomb gauge remains an open problem~\cite{Vandersickel:2012tz}.
Nevertheless, various perturbative or non-perturbative aspects
or applications can be (and have been) addressed, e.g. see references~\cite{Gaigg:1990si, Reinhardt:2017pyr}.
For instance, the Hamiltonian light-front formulation of Quantum Chromodynamics (i.e. the gauge theory
of strong interactions)
is considered to be a promising approach to the problem of determining the field theoretic solutions
which describe hadrons, e.g. see~\cite{Bakker:2013cea} and references therein to the large number of related works.
Canonical quantization in the Coulomb gauge also represents a useful approach to the exploration of confinement in QCD, e.g. see reference~\cite{Reinhardt:2016ogz} for a review.

We note that, for non-Abelian gauge field theory, a
gauge fixing can generally not be realized in a global manner in the space of all gauge fields,
i.e. the so-called \emph{Gribov problem}~\cite{Gribov:1978}
which finds its mathematical expression in a theorem of I.~M.~Singer.
The latter theorem as well as any careful study of finite gauge transformations in non-Abelian gauge theories
rely on the consideration of a specific asymptotic behavior of gauge fields
in order to render the configuration space
mathematically precise: this rules out some gauge choices like the axial gauge~\cite{DeWitt:2003pm}.
Yet, the Gribov problem is related to large gauge transformations, i.e. non-perturbative calculations,
see~\cite{Sobreiro:2005ec,Vandersickel:2012tz,Lechtenfeld:2013faa} and references therein for recent reviews.

%%%%%%%%%%%%%%%%%%%%%%%%%%%%%%%%%%%%%%%%%%%%%%%%%%%%%%%%%%%%%%%%%%%%%%%%%%%%%%%%%%%%%%%%%%%
\subsection{On the Gupta-Bleuler approach}
The quantization procedure of Gupta and Bleuler cannot be applied in the non-Abelian case
since the addition of a term $-\frac{1}{2} \, \Tr \, (\pa_\m A^\m )^2$ to the gauge invariant YM Lagrangian
yields the modified YM equation $0 = D_\m F^{\m \n} + \pa^\n (\pa \cdot A)$:
application of $\pa_\n$ leads to $\Box (\pa \cdot A) = - \ri q \pa_\n [A_\m , F^{\m \n} ]$,
i.e. the fields $\pa^\m A^a _\m$ are not free fields.
This implies~\cite{Das:2008,Kugo}
that one cannot decompose them in a time invariant manner
into positive and negative frequency parts so as to impose a  subsidiary condition
of the form $\pa^\m A^{(+)} _\m (x) \, |\Psi \rangle =0$
holding for all times. Thus, the subsidiary conditions of Gupta-Bleuler and of Nakanishi-Lautrup
are not consistent with time evolution in the non-Abelian case.

%%%%%%%%%%%%%%%%%%%%%%%%%%%%%%%%%%%%%%%%%%%%%%%%%%%%%%%%%%%%%%%%%%%%%%%%%%%%%%%%%%%%%%%%%%%
\subsection{Path integral quantization}
Instead of the canonical quantization, we can consider Feynman's path integral approach
to the quantization of gauge field theories.
In this framework, one functionally integrates over all gauge fields $A^\m $, i.e. one has a functional integral of the form
$\int {\cal D}A \, \re^{\frac{\ri}{\hbar} S_{\textrm{inv}} [A]}$. However, in the latter integral,
the gauge fields are overcounted since all gauge equivalent fields
should only be counted once. The well-known remedy, put forward by Faddeev and Popov (FP), consists of
 the choice of a gauge fixing slice in the space of all gauge fields and in the introduction of the
corresponding FP determinant in the functional integral.
By introducing FP ghost and anti-ghost fields, both of these contributions can be rewritten in a local form
so that the action in the exponential (over which one integrates  in the functional integral) becomes a
 total action
$S_{\textrm{tot}} \equiv S_{\textrm{inv}} + S_{\textrm{fix}} + S_{\textrm{FP}}$.
For a \emph{generalized gauge fixing condition} of the form $f(A) = B$ (where $f$ and $B$ denote given Lie algebra-valued
 functions and where $B$ does not depend on $A$),    the ghost action depends on the Lie algebra-valued
ghost and anti-ghost fields
$c, \bar c$ and it has the structure $ S_{\textrm{FP}} = \int d^4 x \int d^4y \, \bar c ^a (x) {\cal M}_{ab} (x,y) \, c^b (y)$
where
\[
{\cal M}_{ab} (x,y) \equiv \left. \frac{\delta f^a ( \, ^{\omega} \!\! A (x))}{\delta \omega ^b (y)} \right|_{\omega =0}
\qquad (\, \mbox{with} \ \; ^{\omega} \!\! A_\m \equiv A_\m + D_\m \omega \, )
\, .
\]
For instance, for the \emph{homogeneous Coulomb gauge} $\Div \vec A =0$, we have
\begin{align}
{\cal M}^{ab} (x,y)= \pa_x ^i  D^{ab}_{x^i} \, \delta (x-y) \, , \qquad
\mbox{hence}  \quad S_{\textrm{FP}} = \int d^4x \, \Tr \, ( \bar c \, \pa^i  D_i c )
\, ,
\label{eq:FPCb}
\end{align}
and for the  \emph{homogeneous special axial gauge} $A^3 =0$, we have
(upon implementation of $A^3=0$)
\begin{align}
{\cal M}^{ab} (x,y)= \delta ^{ab}
\pa_{x^3}   \, \delta (x-y) \, , \qquad
\mbox{hence}  \quad S_{\textrm{FP}} = \int d^4x \, \Tr \, ( \bar c \, \pa_{3} c )
\, .
\label{eq:FPaxG}
\end{align}

The field-dependent derivative $ \pa^i  D_i $ in the action~\eqref{eq:FPCb} corresponds to the
field-dependent term in the Dirac bracket~\eqref{eq:YMradDir} and gives rise to ghost loops
in the Coulomb gauge, e.g. see reference~\cite{Leibbrandt:1994}.
The absence of a field-dependent term in the axial gauge FP action~\eqref{eq:FPaxG}  reflects the absence of such terms
in the Dirac brackets of $A^1, A^2, \pi^1, \pi^2$ and implies that the FP-ghosts decouple
in the special axial gauge, such gauges being referred to as `ghost-free'
or `physical gauges' (see however~\cite{Heinzl:1995jn} and references therein for subtleties related to infrared divergences):
This is convenient, but these gauges  also come along with a number of complications,
e.g. see references~\cite{Bassetto:1991ue,Leibbrandt:1994,Schweda-book:1998,Gaigg:1990si} for a general discussion and assessment.

%%%%%%%%%%%%%%%%%%%%%%%%%%%%%%%%%%%%%%%%%%%%%%%%%%%%%%%%%%%%%%%%%%%%%%%%%%%%%%%%%%%%%%%%%%%
\subsection{BRST quantization}

If one considers the Lorenz gauge, then the total action  for pure YM theory appearing in the path integral
over $A^\m$ has the form
\begin{align}
S_{\textrm{tot}}  \equiv
S_{\textrm{inv}} + S_{\textrm{fix}} + S_{\textrm{FP}}
\equiv
\int d^4 x \, \textrm{Tr} \, \Big[
-\frac{1}{4} \, F^{\m \n} F_{\m \n} + b \, (\pa_\m A^\m ) + \frac{\xi}{2} \, b^2 + \bar c \, \pa^\m D_\m c
\Big]
\, ,
\label{eq:TotAct}
\end{align}
where the auxiliary field $b$ is a real Lie algebra-valued scalar field and $\xi$ a gauge parameter.
By construction this action is not gauge invariant, but it is invariant
under the so-called \emph{BRST transformations} (Becchi, Rouet, Stora 1974~\cite{\becchi}, Tyutin 1975~\cite{Tyutin:1975qk}).
The latter define a global symmetry (parametrized by a constant, anticommuting parameter)
and represent a relic of local gauge symmetry. By virtue of Noether's first theorem, this invariance
of the gauge fixed action functional $S_{\textrm{tot}}$
yields a conserved charge, the so-called BRST charge.

More precisely~\cite{Kugo,Nakanishi:1990qm,Das:2008},
in the \textbf{Abelian case,}
the gauge fixing action~\eqref{eq:TotAct} leads to a \emph{BRST charge}
of the form $Q= \int d^3x \, \big[ F^{0i} \pa_i c - b \dot{c} \big]$.
If one uses the equation of motion $\pa_\m F^{\m \n} = \pa^\n b$ of the gauge field, then the expression for $Q$ reduces to
\begin{align}
\label{eq:BRSTcharge}
Q= - \int d^3x \,  b \stackrel{\leftrightarrow}{\pa_0} c
\, .
\end{align}
The total action~\eqref{eq:TotAct} is also invariant under the rescaling of ghosts
$c \mapsto \textrm{e}^{\rho} c, \, \bar c \mapsto  \textrm{e}^{-\rho} \bar c$
(with a constant parameter $\rho$)
which leads to a conserved \emph{ghost number charge}
\begin{align}
Q_c= - \int d^3x \,  \bar c \stackrel{\leftrightarrow}{\pa_0} c
\, ,
\end{align}
In the Hamiltonian (canonical) formulation of
quantum theory, one requires  that the physical states $|\Psi \rangle$
are invariant under both operators $Q$ and $Q_c$, i.e. the
 \begin{align}
 \mbox{\textbf{Kugo-Ojima subsidiary condition:}}
 \qquad
 Q \, |\Psi \rangle =0
\, , \qquad
 Q_c \, |\Psi \rangle =0
 \, .
 \label{eq:KOsc}
 \end{align}
For the Fourier components of the fields $c,b$, this condition implies
\begin{align}
\label{eq:BRSTinvStates}
c^{(+)} (\vec k) \, |\Psi \rangle = 0\, , \qquad
b^{(+)} (\vec k) \, |\Psi \rangle = 0
 \qquad \mbox{for all} \ \; \vec k
\, .
\end{align}
According to the first relation,
the states do not involve ghost particles. By virtue of the second relation and  the equation of motion of $b$
(i.e. $b = -\frac{1}{\xi } \, \pa^\m A_\m$), these states are annihilated by $k^\m A_\m ^{(+)} (\vec k)$,
i.e. the Gupta-Bleuler subsidiary condition~\eqref{eq:GBsc} written in momentum space.
We note that for the Abelian theory, the last term in~\eqref{eq:TotAct}
does not involve a coupling to the gauge field, hence the ghost fields decouple in this case:
\emph{For Abelian gauge theory, the BRST approach then amounts to an elegant formulation of the Gupta-Bleuler method
in which the BRST symmetry allows us to eliminate the unphysical degrees of freedom.}

For the case of \textbf{non-Abelian YM-theories}, where the Gupta-Bleuler method
no longer works, the BRST quantization method can be applied straightforwardly.
Actually, this approach to quantization can be applied to quite general field theories involving local symmetries
and it can be used to implement quite general
linear or non-linear gauge fixing conditions, e.g. see~\cite{Piguet:1995,Schweda-book:1998}
and references therein for the Lagrangian framework and~\cite{Henneaux:1992, Rothe:2010} for the  Hamiltonian framework.
Concerning the EMT which we discussed in the previous sections, we note that
it does not only receive contributions from the gauge invariant YM action, but also from the gauge fixing and ghost
terms -- see expression~\eqref{eq:TotAct} for a Lorentz covariant gauge fixing.
However, the latter terms in the action represent a BRST-exact functional, i.e.
$ S_{\textrm{fix}} + S_{\textrm{FP}}$ has the form of a graded commutator of the BRST charge $Q$ with a gauge fixing fermion
$\Phi_{\textrm{gf}} $:
$S_{\textrm{fix}} + S_{\textrm{FP}} = [Q, \Phi_{\textrm{gf}} ]$.
This implies that their contribution $T^{\m \n}_{\textrm{gf}}$ to the total EMT is also BRST-exact\footnote{In this respect,
we note that the EMT can equivalently be defined (e.g. see references~\cite{Forger:2003ut, Blaschke:2016ohs})
by coupling the system to an external gravitational field described by a fixed metric tensor field $(g_{\m \n } (x) )$
that is BRST-invariant: the EMT in Minkowski space is then given by the flat space limit,
i.e. $g_{\m \n } (x) = \eta_{\m \n}$, of the Einstein-Hilbert EMT in curved space as defined by
$T^{\mu \nu}  [\vp, \mg] \equiv
\frac{-2}{\sqrt{|g|}}
\,  \frac{\delta S_M [\vp , \mg ] }{\delta g_{\mu \nu} }$
where  $\mg \equiv  (g_{\m \n})$ and $g \equiv \textrm{det} \, \mg $.}
and thereby BRST invariant by virtue of the nilpotency of the BRST operator.
This ensures that the matrix elements of the operator $: \!  T^{\m \n}_{\textrm{gf}}\! \! :$ between physical states
$|\Psi \rangle, |\Psi ' \rangle$ vanishes~\cite{Leader:2013jra} due to the subsidiary condition~\eqref{eq:KOsc}.

%%%%%%%%%%%%%%%%%%%%%%%%%%%%%%%%%%%%%%%%%%%%%%%%%%%%%%%%%%%%%%%%%%%%%%%%%%%
\section{Matter field  interacting with a gauge field}
\label{sec:matter}

For simplicity, we consider the case of a complex \emph{scalar field} $\phi$ of charge $e$
in $\br^n$ which is minimally coupled to an \emph{Abelian gauge field} $(A^\mu )$.
The  matter field Lagrangian then reads
\begin{align}
{\cal L}_M
(\phi ,A) = (D^\m \phi^*) (D_\m \phi) - m^2 \phi^* \phi
\, , \qquad \mbox{with} \ \;
D_\mu \phi \equiv \pa_\mu \phi + \ri e A_\mu \phi \, , \ \;
D_\mu \phi^* \equiv (D_\mu \phi )^*
\, .
\end{align}
The complete action $S[A, \phi ] \equiv S_{\textrm{gauge}} [A] + S_M [\phi, A]$
now yields the Maxwell equation $\pa_\n F^{\n \m} = j^\m$ where
\begin{align}
j^\m \equiv  j^\mu (\phi , A ) \equiv \ri e \left[  \phi^* D^\mu \phi
- \phi D^{\mu } \phi^*  \right]
\, ,
\label{eq:mattercurrent}
\end{align}
represents the matter current.

The EMT for the minimally coupled field $\phi$ has~\cite{Blaschke:2016ohs} the  expression
%\eqref{eq:ImpScalEMT}
\begin{align}
\label{eq:ImpScalEMT}
T_{\txt{int}}^{\mu \nu} (\phi, A) \equiv
T_{M, \txt{can}}^{\mu \nu} - j^\m A^\n
= (D^{\mu} \phi^* ) ( D^\n \phi)
+ (D^{\mu} \phi ) ( D^\n \phi^*) - \eta^{\m \n} {\cal L}_M (\phi ,A)
\, ,
\end{align}
where
$T_{M, \txt{can}}^{\mu \nu} \equiv \frac{\pa {\cal L}_M}{\pa(\pa_\m \phi)} \, \pa^\n \phi + \frac{\pa {\cal L}_M}{\pa(\pa_\m \phi ^*)} \, \pa^\n \phi ^*
- \eta^{\m \n} \, {\cal L}_M$
is the canonical EMT of the matter field.
The canonically conjugate momenta associated
to $\phi^*$ and $\phi$ are given by the covariant derivatives of the fields:
$\pi \equiv {\pa {\cal L}}/{\pa \dot{\phi}^*} = D_0 \phi$
and $\pi^* \equiv {\pa {\cal L}}/{\pa \dot{\phi}}= D_0 \phi^*$.
Thus the components
$P^\n_{\txt{int}} \equiv \int d^{n-1}x \, T_{\txt{int}}^{0 \nu} [\phi, A]$ of the
energy-momentum vector take the form
\begin{align}
\label{eq:Pintscalar}
P^0_{\txt{int}} &=  \int d^{n-1}x \; \left[ \pi^* \pi + (\vec{D} \phi^*)(\vec D \phi)  + m^2 \phi^* \phi \right]
\, , \nn\\
P^k_{\txt{int}}  &=
- \int d^{n-1}\! x \ \left[ \pi^* \, D_k \phi + \pi \, D_k \phi^* \right]
\, .
\end{align}
By construction these expressions are gauge invariant.

Given the minimal coupling of matter fields, it does not come as a surprise that the charges $P^\m_{\txt{int}}$
generate gauge covariant translations,
\begin{align}
 \Boxed{
\{\vp (x) , P^\m_{\txt{int}} \} = (D^\m \vp)(x)
}
\qquad \mbox{for} \ \; \vp \in \{ \phi, \phi^*, \pi, \pi^* \}
\, ,
\label{eq:PgenCovTransl}
\end{align}
and that they satisfy a non-Abelian algebra involving the field strength tensor of the gauge field:
\begin{align}
 \Boxed{
 \{ P^\m_{\txt{int}} , P^\n_{\txt{int}} \}  = - \int d^{n-1} x \, F^{\m \n} j^0
}
\qquad \mbox{with}
\ \
j^0 = \ri e ( \phi^* \pi - \phi \pi^* )
\, .
\label{eq:NonAbelianAlg}
\end{align}
%Here, $j^0 [\phi, A]$ represents the time component of the matter current %\eqref{eq:mattercurrent}

The \emph{kinematical energy-momentum vector} $P^\m_{\txt{kin}}$ of matter which generates ordinary space-time translations is presently defined by
\begin{align}
 \Boxed{
 P^\m_{\txt{kin}} \equiv P^\m_{\txt{int}}  + \int d^{n-1} x \, A^{\m} j^0
}
\label{eq:kinscalEMV}
\end{align}
and it satisfies
\begin{align}
 \{\vp , P^\m_{\txt{kin}}  \}  = \pa^\m \vp \,, \qquad
 \{ P^\m_{\txt{kin}} , P^\n_{\txt{kin}} \} = 0
\, .
\label{eq:PropKinEMV}
\end{align}
In fact, by comparing the redefinition~\eqref{eq:kinscalEMV} with relation~\eqref{eq:ImpScalEMT} we conclude that $P^\m_{\txt{kin}}$
is nothing else but the canonical energy-momentum vector of the free scalar field,
\begin{align}
\Boxed{
P^\m_{\txt{kin}} = P^\m_{\txt{can}}
 }
\, ,
\end{align}
which was of course to be expected.
The four-vectors $(P^\m_{\txt{int}})$ and
$(P^\m_{\txt{kin}}) = (P^\m_{\txt{can}})$
of a matter field which is minimally coupled to a
gauge field $(A^\m)$ can be  compared to the vectors
%$\vec p_{\txt{kin}} \equiv
$ m\dot{\vec x} = \vec p - e \vec A $ and
%$\vec  p_{\txt{can}} \equiv
$\vec p$ for a charged particle which is minimally
coupled to a vector potential $\vec A$ in classical mechanics.
The angular momentum for a scalar or Dirac field can also be discussed along the previous lines:
for different expressions and aspects, we refer to~\cite{Leader:2013jra, Wakamatsu:2014zza}.

%%%%%%%%%%%%%%%%%%%%%%%%%%%%%%%%%%%%%%%%%%%%%%%%%%%%%%%%%%%%%%%%%%%%%%%%%%%%
%%%%%%%%%%%%%%%%%%%%%%%%%%%%%%%%%%%%%%%%%%%%%%%%%%%%%%%%%%%%%%%%%%%%%%%%%%%%
\section{On the observables of (angular) momentum in gauge theories}\label{sec:obervables}

As in our previous treatment of gauge theories, we again consider the four dimensional case.

From the physical point of view, the components $p^i \equiv T^{0i}$ of the EMT represent the \emph{density $\vec p$
of linear momentum} while the components $(M^{0jk})$ of the angular momentum tensor represent the \emph{density of total angular
momentum,} i.e. of orbital angular momentum $\vec l$ and of intrinsic (spin) angular momentum $\vec s$:
\begin{align}
\vec P \equiv \int_{\br ^3 } d^3x \, \vec p \, , \qquad
\vec J \equiv \int_{\br ^3 } d^3x \, \vec j \equiv  \int_{\br ^3 } d^3x \, ( \vec l + \vec s \, )
\equiv \vec L + \vec S \, .
\end{align}
Any two densities differing by a superpotential term, e.g.
$p^i \equiv T^{0i}$ and $(p^i )^{\prime}= (T^{0i} )^{\prime}= T^{0i} + \pa_j \chi ^{0ji}$
(where $\chi ^{0ji}$ decreases fast enough at spatial infinity)
yield the same integrals, i.e. charges $P^i$ (and similarly for $J^i , \, L^i$ and $ S^i$).
In quantum field theory, the latter charges become self-adjoint operators which play an important role, e.g.
in characterizing the physical states (momentum, spin or helicity).
\emph{Two classically equivalent charges may eventually give rise  to operators
in quantum theory which have quite different properties, e.g. satisfy different commutation relations.}
 These issues have physical consequences
in quantum electrodynamics for instance for the characteristics of laser beams or in quantum chromodynamics for instance for the spin of the nucleon,
the latter being made up of the angular momenta of its constituents (quarks and
gluons)~\cite{Leader:2013jra, Wakamatsu:2014zza, Leader:2015vwa}.
In view of these physical applications,
we briefly summarize here the naturally given classical expressions for the densities of momentum and angular momentum
of a gauge field encountered in section~\ref{sec:LagrForm}, as well as some of the expressions put forward in the
literature~\cite{Leader:2013jra, Wakamatsu:2014zza, Leader:2015vwa}.
We refer to the latter work as well as
to~\cite{Steinmann:2000nr,Lowdon:2014dda, Wakamatsu:2019ain, Lowdon:2020gsj}
for a discussion of problems related to quantization, in particular
the issue of gauge transformations of operators.

The \emph{canonical expressions}~\eqref{eq:canEMTYM}, \eqref{eq:CanAMTgf}
for the EMT and angular momentum tensor of a gauge field
yield  \emph{gauge-dependent densities} for the linear and angular momentum:
The latter can be read off from expressions~\eqref{eq:canonP} and~\eqref{eq:CanJLS},
\begin{align}
\vec p _{{\txt{can}}}=  \Tr \,  (E_{i\, } \vec{\nabla} A^i )
\, , \qquad
\vec l _{{\txt{can}}}=  \Tr \,  \big[ E_{i\, } (\vec x \times \vec{\nabla})  A^i \big]
\, , \qquad
\vec s _{{\txt{can}}}=  \Tr \,  (\vec E \times \vec A)
\, .
\label{eq:CanDensPLS}
\end{align}
In section~\ref{sec:HamFormMaxwell} we saw that, within the extended
Hamiltonian formalism, the conditions~\eqref{eq:gfcond0} only fix the Lagrange multipliers, but leave the gauge field unfixed.
The fundamental Poisson brackets for $A^\m$ and its canonically conjugate momentum
$\pi_\m$
(which hold at fixed time $t$) then
imply standard Poisson brackets for the components
of the spin momentum $\vec S _{{\txt{can}}} \equiv \int_{\br ^3 } d^3x \, \vec s _{{\txt{can}}}$,
i.e. the Poisson algebra relations~\eqref{eq:BracketSS}.
%\begin{equation}
%\label{eq:Spoissbra}
%\{ S_{{\txt{can},i}}, S_{{\txt{can},j}} \} = \varepsilon_{ijk} S_{{\txt{can},k}} \, .
%\end{equation}
Upon replacing the Poisson bracket by  $1/\ri \hbar$ times the commutator of operators,
we obtain the standard commutator algebra of angular momentum in quantum theory
for $\vec{S} _{{\txt{can}}}, \, \vec{L} _{{\txt{can}}}$ and $\vec{J} _{{\txt{can}}}$.

The so-called \emph{improved expressions} for the EMT and angular momentum tensor of a gauge field
as defined by expressions~\eqref{eq:impEMTYM} and~\eqref{eq:MaxAMT}, respectively, give rise to
\emph{gauge invariant densities} which can be read of from~\eqref{eq:gaugeinvP} and~\eqref{eq:GenLorTr}:
\begin{align}
\label{eq:Belimpdens}
\vec p _{\txt{inv}} =  \Tr \,  ( \vec E \times \vec B )
\,, \qquad
\vec j _{\txt{inv}} =  \Tr \,  \big[ \vec x \times (\vec E \times \vec B) \big]
\, .
\end{align}
For an Abelian gauge field these are the familiar expressions of classical electrodynamics~\cite{Jackson:1998}.
In the sequel we focus on this particular case.

Equivalent expressions of physical interest
can be obtained by a decomposition of the vector field $\vec A$ into its transverse and longitudinal
components. In this respect we recall~\cite{GriffithsBook} that any vector field $\vec A$ which decreases for $|\vec x|\to \infty$
faster than $1/|\vec x|$ admits a unique \emph{Helmholtz decomposition}
\begin{align}
\label{eq:Helmholtz}
\Boxed{
\vec A = \vec A_{\|} + \vec A_{\perp} = - \vec{\nabla} V + \vec{\nabla} \times \vec C
}
\, , \qquad \txt{where} \ \; \vec{\nabla} \times \vec A_{\|} = \vec 0 \, , \quad
\vec{\nabla} \cdot \vec A_{\perp} = 0
\, ,
\end{align}
and
\[
V(\vec x \, ) = \frac{1}{4 \pi} \int_{\br ^3} d^3x' \, \frac{\vec{\nabla} ^{\prime} \cdot \vec A (\vec x ^{\, \prime})}{|\vec x - \vec x ^{\, \prime}|}
\, , \qquad
\vec C (\vec x \, ) = \frac{1}{4 \pi} \int_{\br ^3} d^3x' \, \frac{\vec{\nabla} ^{\prime} \times \vec A (\vec x ^{\, \prime})}{|\vec x - \vec x ^{\, \prime}|}
\, .
\]
In Fourier space, the transversality and longitudinality conditions become $\vec k \cdot \tilde{\vec{A}} _{\perp} (\vec k , t)=0$
and $\vec k \times \tilde{\vec{A}} _{\|} (\vec k , t)= \vec 0$.
It should be noted that the expressions for $ \vec A_{\|}$ and $\vec A_{\perp}$
are non-local in $\vec A$.
For a gauge transformation $\vec A \leadsto \vec A' = \vec A + \vec{\nabla} \alpha$,
we have $\vec A ' _{\perp} = \vec A_{\perp}$ and $\vec A ' _{\|} = \vec A_{\|} + \vec{\nabla} \alpha$, i.e.
$ \vec A_{\perp}$ is gauge invariant.
 We remark that for the more general case of non-Abelian gauge fields, the geometric structure underlying the decomposition~\eqref{eq:Helmholtz}
 is related to the so-called
 \emph{dressing field method} to construct gauge invariants, see~\cite{Francois:2014bya} and references therein.

Using the free field equation $\vec{\nabla} \cdot \vec E =0$, one can easily verify that
the canonical  densities~\eqref{eq:CanDensPLS}
are related by a divergence (superpotential term) to the following \emph{densities considered by Chen, Lu, Sun, Wang, and Goldman}~\cite{Chen:2008ag}:
\begin{align}
\vec p _{{\txt{Chen}}}&= E_{i\,} \vec{\nabla} A^i_{\perp}
\, , &
\vec j _{{\txt{Chen}}}&=\vec l _{{\txt{Chen}}} + \vec s _{{\txt{Chen}}}
\,, &&  \txt{with} \ \;
\left\{
\begin{array}{l}
\vec l _{{\txt{Chen}}}= E_{i\,} (\vec x \times \vec{\nabla})  A^i_{\perp}
\,,\\
\vec s _{{\txt{Chen}}}= \vec E \times \vec A_{\perp}
\, .
\end{array}
\right.
\label{eq:Chendens}
\end{align}

Since the densities~\eqref{eq:Chendens} are gauge invariant, one may as well spell
them out in a convenient gauge, for instance in
the \emph{radiation gauge}, i.e. for gauge potentials $(A^\m )$ satisfying $\vec{\nabla} \cdot \vec A =0$ and $A^0 =0$. Then we have
$\vec E_{\|} =  - \vec{\nabla} A^0=  \vec 0$, hence
$\vec E = \vec E_{\perp} = -\dot{\vec A}   $
(with $\vec{\nabla} \cdot \dot{\vec A}  =0$),
and thereby we obtain the so-called \emph{gauge invariant canonical expressions}~\cite{Leader:2015vwa}
which only involve $\vec E_{\perp}$ and $\vec A _{\perp}$:
\begin{align}
\vec p _{{\txt{gic}}}&= E_{\perp i\, } \vec{\nabla} A^i_{\perp}
\,, &
\vec j _{{\txt{gic}}}&=\vec l _{{\txt{gic}}} + \vec s _{{\txt{gic}}}
\,, && \txt{with} \ \;
\left\{
\begin{array}{l}
\vec l _{{\txt{gic}}}= E_{\perp i\,} (\vec x \times \vec{\nabla})  A^i_{\perp}
\,,\\
\vec s _{{\txt{gic}}}= \vec E_{\perp} \times \vec A_{\perp}
\, .
\end{array}
\right.
\label{eq:gicDens}
\end{align}
(For instance, for plane waves with frequency $\omega$, we have
$\vec A_{\perp} (t, \vec x \,) = \vec A_{\perp 0} ( \vec x \,)\,  \re^{- \ri \omega t}$ with $\vec{\nabla} \cdot {\vec A}_{\perp}  =0$, which implies
$\vec E = \vec E_{\perp} = -\dot{\vec A}_{\perp} = \ri \omega   \vec A_{\perp} $, hence $\vec A_{\perp} =  \frac{1}{\ri \omega} \, \vec E $
is a local field.)
In the present setting (where $\vec E_{\|}= \vec 0$ and $\vec E = \vec E_{\perp}=-\dot{\vec A}$),
the  Poisson brackets are to be chosen to have the Dirac form~\eqref{diracbraA} with $\vec{\pi} = \vec E = - \dot{\vec A} $.
%\begin{equation}
%\label{eq:fundpoissbrack}
%\{ A_i (t, \vec x \, ) , \dot{A}_j (t , \vec y \, ) \} = \delta_{ij}^{\perp} (\vec x - \vec y \, )
%\, , \qquad \txt{where} \ \; \delta_{ij}^{\perp}(\vec x \, ) = \int_{\br ^3 } \frac{d^3k}{(2 \pi)^3}
%\, \re^{\ri \vec k \cdot \vec x} \,
%\big( \delta_{ij} - \frac{k_i k_j}{\vec k ^{\, 2}} \big) \, .
%\end{equation}
 In this case, $\vec L _{{\txt{gic}}}$ and  $\vec S _{{\txt{gic}}}$ represent physically measurable quantities,
 but they cannot really be interpreted as the orbital and spin angular momentum of the electromagnetic field
 due to the fact that they do not satisfy the algebra
 of angular momenta, e.g. the components of the vectorial operator $\vec S _{{\txt{gic}}}$ commute with each other
 --- see~\cite{Wakamatsu:2014zza, Leader:2015vwa}
 and references therein.
 For a recent assessment of the physical issues in the absence or presence of matter and in particular
 the role of boundary terms, we refer to~\cite{Wakamatsu:2019ain}.

%\section{Concluding remarks}

%%%%%%%%%%%%%%%%%%%%%%%%%%%%%%%%%%%%%%%%%%%%%%%%%%%%%%%%%%%%%%%%%%%%%%%%%%%%%%%%%%%%%%%%%%%%%%%%%%%
%%%%%%%%%%%%%%%%%%%%%%%%%%%%%%%%%%%%%%%%%%%%%%%%%%%%%%%%%%%%%%%%%%%%%%%%%%%%%%%%%%%%%%%%%%%%%%%%%%%
%%%%%%%%%%%%%%%%%%%%%%%%%%%%%%%%%%%%%%%%%%%%%%%%%%%%%%%%%%%%%%%%%%%%%%%%%%%%%%%%%%%%%%%%%%%%%%%%%%%
\section{Covariant Hamiltonian approaches}

We recall that our starting point for relativistic gauge field theories was the Lagrangian
formulation -- see~\secref{sec:LagrForm}.
Thereafter, we considered the standard Hamiltonian approach to these theories.
Since time derivatives of fields are treated differently from spatial derivatives in the latter
approach, Lorentz covariance is not manifest.
For this reason, covariant canonical formulations have been sought for which retain as much as possible
the advantages of the standard Hamiltonian approach.
Several such approaches have attracted a lot of attention
during the last decades.
We mention the \emph{multisymplectic approach} following ideas put forward, in particular, towards 1970
by the Warsaw school (notably J.~Kijowski~\cite{\Kijowski},
K.~Gaw\c{e}dzki~\cite{Gawedzki:1972} and W.~M.~Tulczyjew~\cite{Kijowski1979})
and independently by the Spanish school~\cite{\GarciaPerez, Carinena86}
 as well as H.~Goldschmidt and S.~Sternberg~\cite{Goldschmidt73}: for this set-up
there exist numerous variants, e.g. see reference~\cite{Roman-Roy2009} for a partial overview.
 Another formulation is the \emph{covariant phase approach}
 based on the  so-called covariant phase space, i.e. the infinite-dimensional space
 of all solutions of the field equations. For this set-up, one can adopt the view-point
 of \emph{symplectic geometry} (following again the Warsaw school as well as more recent work of E.~Witten~\cite{\WittenEtal}
 and G.~Zuckerman~\cite{Zuckerman}) or consider the so-called \emph{Peierls bracket} introduced
 by R.~E.~Peierls~\cite{Peierls:1952cb} and thoroughly investigated by B.~DeWitt~\cite{DeWitt:2003pm}.
 There exist relationships between all of these approaches as
 well as the standard Hamiltonian approach that we followed here
 (e.g. see references~\cite{Barnich:1991tc, Forger:2003jm,  Khavkine:2014kya, Forger:2015tka}
  for some results
 in this direction), all formulations having their advantages and shortcomings.
 Since the covariant approaches rely on physical and mathematical concepts that are sensibly different
 from the ones of  the standard Hamiltonian approach, we will not expand further on these issues here
 and rather defer this discussion (in particular the treatment of symmetries and conserved currents/charges
in gauge field theories) to a separate work.

% \vskip 1.2truecm
%%%%%%%%%%%%%%%%%%%%%%%%%%%%%%%%%%%%%%%%
\subsection*{Acknowledgments}

We are indebted to M\'eril Reboud for collaboration in an early stage of this work.
F.~G. wishes to thank  F.~Delduc and K.~Gaw\k{e}dzki for helpful comments.

%\newpage
%%%%%%%%%%%%%%%%%%%%%%%%%%%%%%%%%%%%%%%%%%%%%%%%%%%%%%%%%%%%%%%%%%%%%%%%%%%%%%%%%%%%%%%%%%%%%%%%%%%%%%%%%%%%%%%%%%%%%
%%%%%%%%%%%%%%%%%%%%%%%%%%%%%%%%%%%%%%%%%%%%%%%%%%%%%%%%%%%%%%%%%%%%%%%%%%%%%%%%%%%%%%%%%%%%%%%%%%%%%%%%%%%%%%%%%%%%%
\appendix

%%%%%%%%%%%%%%%%%%%%%%%%%%%%%%%%%%%%%%%%%%%%%%%%%%%%%%%%%%%%%%%%%%%%%%%%%%%%%%%%%%%%%%%%%%%%%%%%%%%%%%%%%%%%%%%%%%%%%
\section{Derivation of the gauge invariant currents associated to conformal invariance}\label{sec:Derivation}

In the last paragraph of this appendix, we present a concise and straightforward derivation of the gauge invariant currents
associated to the conformal symmetry
(EMT, angular momentum tensor, scale current,...) as well as of their conservation laws, this derivation
providing also the superpotential terms which relate these currents to the canonical expressions.
Our argumentation generalizes the one considered for free Maxwell theory in four dimensions
by the author of reference~\cite{Nair:2005iw}.
It relies on the use of the well-known conformal Killing vector fields of Minkowski
space-time and on a gauge covariantization procedure, i.e. expressing ordinary derivatives in terms of gauge covariant derivatives.
The latter procedure has been repeatedly (re-)discovered in the literature (in particular in the case of translation invariance),
one of the earliest (if not the first) consideration being due to R.~Jackiw~\cite{Jackiw78}.
Our treatment of conformal invariance of pure YM theories applies in $\br ^4$
and more generally in $\br ^n$ for the particular case of Poincar\'e transformations.
Although the considered geometric approach~\cite{Nair:2005iw} is appealing
and works quite well in the case of four dimensions, it should be noted that
the description of infinitesimal conformal transformations by Lie derivatives
requires some modifications in a space-time of arbitrary dimension: we will first elaborate
on this fact while generalizing some results of reference~\cite{Jackiw:2011vz}.

%%%%%%%%%%%%%%%%%%%%%%%%%%%%%%%%%%%%%%%%%%%%%%%%%%%%%%%%%%%%%%%%%%%%%%%%%%%%%%%%%%%
\paragraph{Conformal group:}
By definition, conformal transformations in Minkowski space-time $(\br ^n, \eta )$
 are transformations $x\leadsto x' (x)$ which preserve the angles,
i.e. the Minkowski metric is preserved under these transformations up to a scale factor:
$ds^2 \leadsto \re^{\lambda } ds^2$
where $\lambda$ represents a constant real parameter~\cite{Sundermeyer:2014kha}.
The associated infinitesimal transformations $x^{\prime \m} (x)  \simeq x^\m + \xi^\m (x)$
are generated by \emph{conformal Killing vector fields}
$\xi \equiv \xi^\m \pa_\m$, i.e. solutions of the
%\emph{conformal Killing equation}
\begin{align}
\label{eq:CKEq}
\mbox{conformal Killing equation :} \qquad
 \pa_\m \xi_\n + \pa_\n \xi_\m
 - \frac{2}{n} \, (\pa_{\rho} \xi^{\rho} ) \, \eta _{\m \n} =0
 \, .
\end{align}
The general solution of this equation is given by
\begin{align}
\label{eq:CKVF}
\xi_\m  = a_\m + \varep_{\m \n} x^\n
  + \rho \, x_\m + 2 \, (c \cdot x) \, x_\m - c_\m x^2
\, ,
\end{align}
where $a_\m \, , \rho \, ,  c_\m$ and $ \varep_{\m \n}  = - \varep_{\n \m} $ are constant real parameters.
For Poincar\'e transformations we have
$\xi_\m (x) = a_\m + \varep_{\m \n} x^\n$: this transformation also preserves the lengths and solves the
ordinary Killing equation $  \pa_\m \xi_\n + \pa_\n \xi_\m =0$ (whence $\pa_{\rho} \xi^{\rho} =0$).
The parameter $\rho$ labels \emph{scale transformations (dilatations)} $x' = \re^{\rho} x$ while $(c_\m )$
labels \emph{conformal boosts (special conformal transformations).}

%%%%%%%%%%%%%%%%%%%%%%%%%%%%%%%%%%%%%%%%%%%%%%%%%%%%%%%%%%%%%%%%%%%%%%%%%%%%%%%%%%%
\paragraph{Transformation laws of fields:}
The gauge potential $(A_\m )$ and the associated field strength
${F} _{\mu \nu} \equiv \pa_{\mu} {A} _{\nu} - \pa_{\nu}
{A} _{\mu} + \ri \mq \, [ {A} _{\mu}  ,  {A} _{\nu} ]$
introduced in~\secref{sec:GenSetUp} represent relativistic tensor fields and thereby transform
with the Lie derivative $L_\xi$ under infinitesimal diffeomorphisms generated by a vector field
$\xi \equiv \xi^\m \pa_\m$:
\begin{subequations}
\begin{align}
\label{eq:LieDer1}
\delta_\xi A_\m &=  \big( L_\xi A \big)_\m = \xi^\n \pa_\n A_\m + (\pa_\m \xi^\n) A_\n
\, , \\
\label{eq:LieDer2}
\delta_\xi F_{\m \n} &=  \big( L_\xi F \big)_{\m \n} = \xi^\rho \pa_\rho F_{\m \n} + (\pa_\m \xi^\rho) F_{\rho \n}
+ (\pa_\n \xi^\rho) F_{\m \rho}
\, .
\end{align}
 \end{subequations}
 For the particular case of the vector field $\xi  [a, \varepsilon ]= (a^\m + \varep^{\m \n} x_\n )\pa_\m$
 describing infinitesimal Poincar\'e transformations, the variation~\eqref{eq:LieDer1} yields the standard form
 of infinitesimal translations and Lorentz transformations as considered in
 equations~\eqref{eq:TransTransA} and~\eqref{eq:LtransforA}, respectively.
 For the case of scale transformations where $\xi^\m [\rho ] =  \rho \, x^\m$ in $\br^n$,
the variations~\eqref{eq:LieDer1}-\eqref{eq:LieDer2} become
\begin{subequations}
\begin{align}
\label{eq:GeoScale1}
\delta_{\xi [ \rho ]} A_\m  & = \rho \, (x^\n \pa_\n A_\m + 1 \, A_\m )
\\
\label{eq:GeoScale2}
\delta_{\xi [ \rho ]} F_{\m \n} & = \rho \, (x^\n \pa_\n F_{\m \n} + 2 \, F_{\m \n} )
\, .
\end{align}
 \end{subequations}
Here, the first equation  is the standard form of a scale transformation
$\delta_\rho A_\m  $ of a vector field in $\br^4$ with parameter $\rho$ 
(and so is the second equation for a tensor field with components $F_{\m \n}$).
The scale factors in the last terms of the geometric transformations~\eqref{eq:GeoScale1}-\eqref{eq:GeoScale2}
simply reflect the rank of the covariant tensor fields under consideration.
However, for a space-time of generic dimension $n$,
the \emph{scale transformation of a vector field,}
i.e.
\begin{align}
\label{eq:ScaleTrafA}
\delta_{\rho} A_\m  & = \rho \, ( x^\n \pa_\n A_\m + d_A (n) \, A_\m )
\end{align}
 involves the \emph{scale (canonical) dimension} $d_A (n) \equiv \frac{n-2}{2}$ of this field in $\br^n$~\cite{Sundermeyer:2014kha}.
To recover this transformation law of $A_\m$ for $\xi^\m [\rho ] =  \rho \, x^\m$
from the geometric transformation law~\eqref{eq:LieDer1}, the latter
 has to be supplemented by an additional term~\cite{Jackiw:2011vz}:
  \begin{align}
  \label{eq:TraFoAmu}
  \Boxed{
\delta_{\textrm{CG}} A_\m   \equiv  \big( L_\xi A \big)_\m  + \frac{n-4}{2n} \, (\pa_\n \xi^\n ) \, A_\m
}
\, .
\end{align}
For  Poincar\'e transformations, i.e. for $\xi ^\m [a, \varepsilon ]= a^\m + \varep^{\m \n} x_\n $,
the last term in expression~\eqref{eq:TraFoAmu} vanishes.
For dilatations, i.e. for $\xi^\m [\rho ] =  \rho \, x^\m$,
relation~\eqref{eq:TraFoAmu} yields the correct transformation law~\eqref{eq:ScaleTrafA} of $A_\m$.
It also does for conformal boosts, i.e. for  $\xi^\m [c] =  2 \, (c \cdot x) \, x^\m - c^\m x^2$.
Henceforth, the variation~\eqref{eq:TraFoAmu} (whose last term vanishes for $n=4$ as well as for
  Poincar\'e transformations in $\br^n$) describes all infinitesimal transformations of $A_\m$
associated to the conformal group (whence the label CG in the variation~\eqref{eq:TraFoAmu}).
Expression~\eqref{eq:TraFoAmu} induces the following transformation law of the field strength:
  \begin{align}
  \label{eq:TraFoF}
\delta_{\textrm{CG}} F_{\m \n}  =  \big( L_\xi F \big)_{\m \n}  & + \frac{n-4}{2n} \, (\pa_\rho \xi^\rho ) \, F_{\m \n}
\\
& \ + \frac{n-4}{2n} \, \big\{ \pa_\m (\pa_\rho \xi^\rho ) \, A_\n - \pa_\n (\pa_\rho \xi^\rho ) \, A_\m
 +\ri \mq \, (\pa_\rho \xi^\rho ) \, [A_\m , A_\n ] \big\}
\, .
\nn
\end{align}
In the case of an Abelian gauge field theory, this expression reduces to the results given in reference~\cite{Jackiw:2011vz},
e.g. for dilatations we obtain $\delta_{\rho} F_{\m \n}  = \rho \, (x^\lambda \pa_\lambda F_{\m \n} + \frac{n}{2}\,  F_{\m \n})$.
 The $A$-dependent contributions in~\eqref{eq:TraFoF} (which are not gauge covariant) reflect the fact that
 the  field strength does not represent a primary field for $n\neq 4$.

%%%%%%%%%%%%%%%%%%%%%%%%%%%%%%%%%%%%%%%%%%%%%%%%%%%%%%%%%%%%%%%%%%%%%%%%%%%%%%%%%%%
\paragraph{Conformal invariance and associated conservation laws:}
For $n \neq 4$, the coupling constant $\mq$ of YM-theory is dimensionful and thereby the
YM action is only scale invariant for $n=4$ (by contrast to the Abelian  theory
which is scale invariant for all values of $n$).
Accordingly, we can consider the geometric transformation laws~\eqref{eq:LieDer1} -\eqref{eq:LieDer2}
for all conformal Killing vector fields in $\br^4$ as well as for the
conformal Killing vector fields $\xi ^\m [a, \varepsilon ]$ which generate Poincar\'e transformations in $\br ^n$.
The derivative of $(A_\m)$  in the transformation law~\eqref{eq:LieDer1}  can be expressed in terms of the
field strength:
\begin{align}
\label{eq:CoLieDer}
\Boxed{
\delta_\xi A_\m =  \xi^\n F_{\n \m} + D_\m (\xi \cdot A )
}
\, , \qquad \mbox{with} \quad
\xi \cdot A \equiv \xi^\n A_\n
\, .
\end{align}
Here, the first term reflects a covariantization of $\xi^\n \pa_\n A_\m$ and the second term represents an infinitesimal gauge transformation of $A_\m$
with (field dependent) parameter $\xi \cdot A $.
As is well known, (e.g. see reference~\cite{Barnich:2001jy}), such a gauge transformation does not contribute to the Noether charge
since it yields a current which vanishes on-shell up to a superpotential term.
We note that the coordinate transformations~\eqref{eq:CoLieDer} of gauge fields 
are often considered in conjunction with local gauge transformations, in particular in the study 
of BRST symmetries in flat or curved space-time, e.g. see~\cite{Gieres:1988qp} and references therein.   

The transformation law~\eqref{eq:LieDer2}  of $F_{\m \n}$ induces the following variation of the Lagrangian
${\cal L} \equiv -  \frac{1}{4} \, \Tr \, ( {F} ^{\mu \nu} {F} _{\mu \nu} ) $ of pure YM theory:
\begin{align}
\label{eq:VarL}
\delta_\xi {\cal L} =  \pa_\m ( \xi^\m {\cal L})
+ \frac{1}{2} \, \Tr \, \big[ {F} ^{\mu \alpha} {F_\alpha} ^{\n}
\, ( \pa_\m \xi_\n + \pa_\n \xi_\m
 - \frac{1}{2} \, (\pa_{\rho} \xi^{\rho} ) \, \eta _{\m \n} ) \big]
\, .
\end{align}
Here, the trace term vanishes by virtue of the conformal Killing equation~\eqref{eq:CKEq} with $n=4$ and it vanishes for all values $n$
if we limit ourselves to Poincar\'e transformations.
Thus, the action is invariant under the corresponding geometric transformations of the gauge fields.
To recover Noether's first theorem, we multiply the equation of motion function
$\frac{\delta S}{\delta A_\m ^a} = D_\n F_a^{\n \m}$ with the variation $\delta_\xi A_\m ^a$:
substitution of~\eqref{eq:CoLieDer} and use of the Leibniz rule  as well as
of the Bianchi identity
($0= D_\rho F_{\m \n} + $ cyclic permutations of the indices) and of
$[D_\m , D_\n ] (\xi \cdot A ) = \ri \mq \, [ F _{\mu \n}  , \xi \cdot A ]$
then yields the result
\begin{align}
\label{eq:Noether1CT}
0= \Tr \, \Big( \frac{\delta S}{\delta A_\m} \, \delta_\xi A_\m \Big) +\pa_\m j^\m
- \frac{1}{2} \, \Tr \, \big[ {F} ^{\mu \alpha} {F_\alpha} ^{\n}
\, ( \pa_\m \xi_\n + \pa_\n \xi_\m
 - \frac{1}{2} \, (\pa_{\rho} \xi^{\rho} ) \, \eta _{\m \n} ) \big]
\, ,
\end{align}
with
 \begin{align}
\label{eq:ConsCurrCT}
j^\m \equiv T^{\m \n}_{\textrm{inv}} \xi_\n + J^\m
\, , \quad
\Boxed{
T_{\txt{inv}} ^{\m \nu}  \equiv
  \Tr \,  ( F^{\m \rho} {F_{\rho}} ^\n ) - \eta ^{\m \n} {\cal L}
  }
  \, , \quad
  \Boxed{
  J^\m \equiv  \Tr \,  \big[ - F^{\m \n} D_\n (\xi \cdot A ) \big]
  }
\, .
\end{align}
As we noted after~\eqnref{eq:VarL}, the last term in~\eqref{eq:Noether1CT}
vanishes for conformal transformations in $\br^4$  and it vanishes for Poincar\'e transformations in $\br^n$.
For the solutions of the equation of motion $0= \frac{\delta S}{\delta A_\m } = D_\n F^{\n \m}$,
the current density $(j^\m )$ is thus conserved in these cases, (i.e. $\pa_\m j^\m \approx 0$)
and the contribution $J^\m$ to $j^\m $
writes
\begin{align}
\label{eq:OnShellJ}
J^\m \approx \pa_\n  \big[ \Tr \,  ( - F^{\m \n}  A^\rho) \xi_\rho  \big] \equiv \pa_\n B^{\m \n}
\qquad  \mbox{with} \ \; B^{\m \n} = -  B^{\n \m}
\, .
\end{align}
This term represents a superpotential term (whose divergence vanishes identically and which does not contribute
to the Noether charges).
More precisely, for the particular case of \emph{translations,} i.e. for $\xi _\m = a_\m$,
this superpotential term is the one encountered in~\eqnref{eq:impEMTYM}
and so is the gauge invariant EMT $ T^{\m \n}_{\textrm{inv}} $ satisfying the
local conservation law $\pa_\m T^{\m \n}_{\textrm{inv}} \approx 0$:  by virtue of equations~\eqref{eq:ConsCurrCT}-\eqref{eq:OnShellJ}
we altogether have
\begin{align}
\label{eq:SummTransl}
\Boxed{
j^\m \equiv
T^{\m \n}_{\textrm{inv}} a_\n + \Tr \,  \big[ - F^{\m \rho} D_\rho (a_\n A^\n ) \big]
 \approx  \left\{ T^{\m \n}_{\textrm{inv}}  + \pa_\rho  \big[ \Tr \,  ( - F^{\m \rho}  A^\n )   \big] \right\} a_\n =  T^{\m \n}_{\textrm{can}} a_\n
 }
 \, .
\end{align}
Thus, we have recovered the canonical EMT and the gauge invariant EMT as well as the superpotential term 
which relates these conserved currents. 
In the case of \emph{Lorentz transformations,}
i.e. for $\xi_\m = \varepsilon_{\m \n} x^\n$,
we get
\begin{align}
\label{eq:LTconsCurr}
J^\m & \approx \frac{1}{2} \, \varepsilon_{\rho \sigma} \pa_\n \Tr \,   \big[   F^{\m \n}  (x^\rho A^\sigma - x^\sigma A^\rho ) \big]
\, , \\
j^\m & = T^{\m \n}_{\textrm{inv}} \xi_\n = - \frac{1}{2} \, \varepsilon_{\rho \sigma}  M^{\m \rho \sigma}_{\textrm{inv}}
\, , \qquad
\mbox{with} \ \;
M^{\m \rho \sigma}_{\textrm{inv}} \equiv x^\rho  T^{\m \sigma}_{\textrm{inv}}
- x^\sigma  T^{\m \rho}_{\textrm{inv}}
\, ,
\nn
\end{align}
and $\pa_\m M^{\m \rho \sigma}_{\textrm{inv}} \approx 0$.
The previous expressions for the angular momentum tensor $M^{\m \rho \sigma}_{\textrm{inv}} $ and
for the superpotential term $J^\m$ coincide with those in~\eqnref{eq:MaxAMT}.
The local conservation law of $M^{\m \rho \sigma}_{\textrm{inv}} $
reflects the on-shell symmetry of the EMT $ T^{\m \n}_{\textrm{inv}} $
since $\pa_\m M^{\m \rho \sigma}_{\textrm{inv}} \approx T^{\rho \sigma}_{\textrm{inv}} -  T^{\sigma \rho}_{\textrm{inv}}$.
As a matter of fact, the EMT $T_{\txt{inv}} ^{\m \nu}$ given by~\eqref{eq:ConsCurrCT} is symmetric off-shell.
For the \emph{scale transformations,} i.e. for $\xi_\m = \rho \, x_\m$, the current
$j^\m = T^{\m \n}_{\textrm{inv}} \xi_\n  =\rho \,  T^{\m \n}_{\textrm{inv}} x_\n$
represents the \emph{dilatation current} whose local conservation law
expresses the (on-shell) tracelessness of the EMT $T^{\m \n}_{\textrm{inv}}$
in four space-time dimensions:
\[
0 \approx \pa_\m j^\m \approx \rho \, T^{\m}_{\textrm{inv} \, \m}
\qquad \mbox{for} \ \; n=4
\, .
\]
The EMT $T^{\m \n}_{\textrm{inv}}$ given by~\eqref{eq:ConsCurrCT} is actually traceless off-shell
for $n=4$.

To conclude, we note that a completely analogous covariantization procedure
can be applied to minimally coupled matter fields $\varphi$: one writes
\[
\delta _\xi \varphi \equiv \xi^\n \pa_\n \varphi =
\xi^\n D_\n \varphi - \ri \mq \, (\xi \cdot A) \varphi
\, ,
\]
where the last term again describes an infinitesimal (field dependent) 
gauge transformation.

\newpage

%%%%%%%%%%%%%%%%%%%%%%%%%%%%%%%%%%%%%%%%%%%
% \nocite{*}
% \printbibliography
%%%%%%%%%%%%%%%%%%%%%%%%%%%%%%%%%%%%%%%%%%%

\providecommand{\href}[2]{#2}
\end{document}